# Joint Covering Congestion Rents in Multi-area Power Systems Considering Loop Flow Effects

Hao Liu, *Student Member, IEEE*, Ye Guo, *Senior Member, IEEE*, Haitian Liu, *Student Member, IEEE*, and Hongbin Sun, *Fellow, IEEE*

***Abstract*—We consider the problem of how multiple areas should jointly cover congestion rents of internal and tie-lines in an interconnected power system. A key issue of our concern is the loop flow problem, which represents discrepancies between scheduled and actual power flow distributions because electric power does not always flow along the most direct paths of transactions. We employ generalized coordinated transaction scheduling (GCTS) for interchange scheduling, which can eliminate dispatch errors caused by loop flow effects and asymptotically converge to the joint economic dispatch (JED) under ideal assumptions. Subsequently, distributed algorithms are proposed for each area to recover multipliers of the global GCTS model, as well as quantifying and pricing its contribution to line congestions. Thereby, all areas and interface bids can jointly cover congestion rents of internal and tie-lines with their merchandise surpluses and profits. Simulations demonstrate the effectiveness of the proposed approach of LMP recovery and joint covering congestion rents.**

***Index Terms*—loop flow effects, congestion rent, generalized coordinated transaction scheduling, locational marginal price recovery**

## I. Introduction

### A. Motivation

In our efforts towards a carbon-neutral society, power system interconnections will play a more and more important role [1-3]: The geographical distributions of renewable energy and load centers are usually opposite, and we also need wide-area coordination and mutual support to guard against extreme weather conditions when wind or solar power are not available in a certain area [4, 5]. Therefore, market mechanisms for an interconnected multi-area power system are of crucial importance.

Deregulated electricity markets have been extensively studied and have been operating for decades in the U.S. and Europe [6]. In the U.S., the standard design of electricity markets is based on the economic dispatch model and locational marginal prices (LMP). For a single-area electricity market, such a design leads to many attractive properties, with one of them being revenue adequacy: the merchandise surplus collected by each ISO through LMP is equal to the congestion rent, which can adequately support any feasible financial transmission right (FTR) auctions [7], [8].

Unfortunately, such an important property does not automatically hold for an interconnected multi-area power system. Take the two-area system in Fig. 1 as an example, assume one of the two tie-lines is congested and there is no other congestion. It is an open question how should the two areas jointly cover the congestion rent of that tie-line. In the implementation of coordinated transaction scheduling (CTS), New York Independent System Operator (NYISO) and ISO-New England (ISO-NE) split tie-line congestion rent by halves [5], which may not be the best solution under various operation conditions.

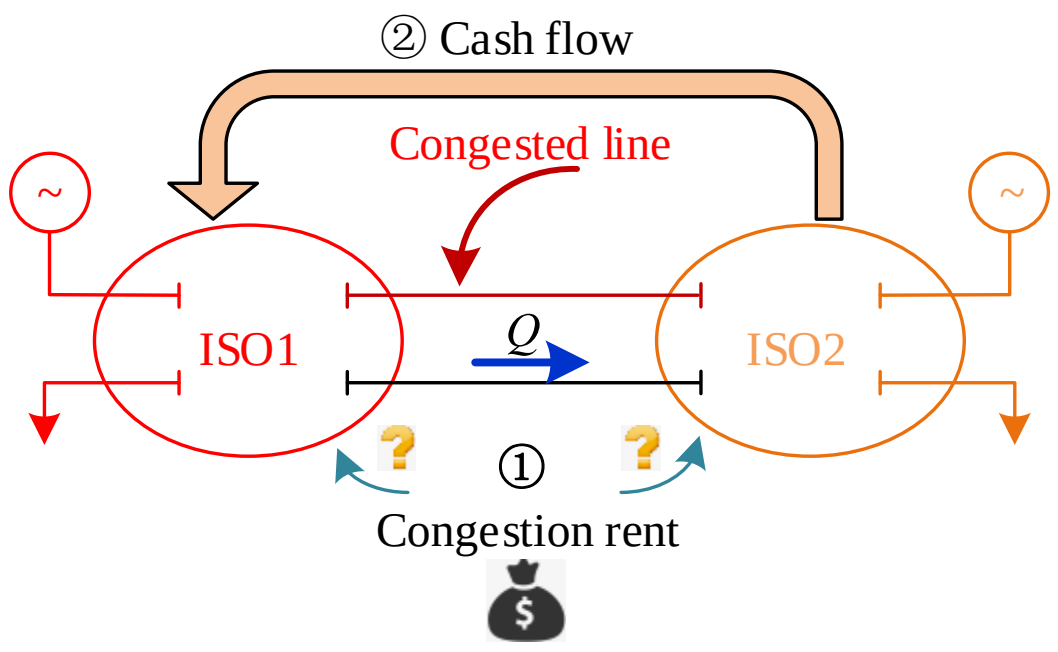

Fig. 1. Interconnected two-area power system with tie-line congested

To answer the question above, we need to decompose the power flow of congested lines into contributions of different kinds of transactions and price each component properly. In particular, not only inter-regional power transactions will flow through tie-lines, a portion of internal transactions in area 1 and 2 will also wheel through the other area physically, contributing positively or negatively to the congestion. Such phenomena are referred to as loop flows, which are power flow along an unintended path that loops away from the most direct geographic path or the contract path [9]. Loop flow problems have already increased economic costs and caused security problems in electricity markets of the U.S. [10], Europe [11], and China [12]. Determining how to model the impact of the loop flow problem is hitherto an open question.

### B. Literature Review

Many procedures are adopted to mitigate the loop flow problem. In the U.S., Southwest Power Pool (SPP) sets phase

The authors are with Tsinghua Berkeley Shenzhen Institute (TBSI), Tsinghua University, Shenzhen, China, 518055. H. Sun is also with the Department of Electrical Engineering, Tsinghua University, Beijing, China, 100084. This work was supported in part by the National Natural Science Foundation of China under Grant 51977115. Corresponding author: Ye Guo, e-mail: guo-ye@sz.tsinghua.edu.cn

angle regulators (PARs) between its border with the California ISO (CAISO) [13]. And in Europe, European Commission tries to solve this problem by dividing pricing zones, optimizing the network topology, and setting PARs [14]. However, these works have not clearly explained how loop flow is generated, thus may not be able to eliminate the loop flow problem.

The occurrence of the loop flow problem is closely related to interchange scheduling between regional markets. Researches on interchange scheduling in interconnected power systems can be divided into two types. First, distributed algorithms to solve the joint economic dispatch (JED) problem have been studied. Lagrange relaxation-based algorithm [15], consensus-based fully distributed algorithm [16], marginal equivalent decomposition algorithm [17], and critical region exploration method [18] have been developed. Traditional distributed algorithms may suffer from slow convergence and computation speed, whereas recent advances have shown much faster convergence [17-19].

Second, novel market mechanisms for interchange scheduling have been studied in multiple ISOs, such as NYISO, ISO-NE, and PJM. Approaches of tie optimization (TO) [5] and coordinated transaction scheduling (CTS) [20] have been developed, but they rely on the proxy model which may lead to substantial errors [21]. Paper [22] puts forward generalized CTS (GCTS), which eliminates the need for proxy buses and can asymptotically converge to JED results under certain conditions. However, in all of these studies, how to jointly cover congestion rent in multi-area power systems has not been adequately addressed.

### C. Contributions

This paper discusses how multiple areas should properly and explicitly cover congestion rents. In particular, contributions of this paper include the following:

i) A distributed algorithm is developed to recover global LMPs. The results are the same as LMPs calculated by the centralized GCTS model.

ii) Loop flow effects are decomposed into different categories. And contributions of each area and interface bids to line congestions are quantified.

iii) Congestion rents are covered proportionally according to the power flow contribution to line congestions.

The rest of this paper is organized as follows: Section II presents preliminary models for the interchange scheduling, including CTS and GCTS. Section III describes the framework of jointly covering congestion rents of both internal lines and tie-lines. Section IV presents a distributed LMP recovery method and its properties. Section V settles each area and interface bids and jointly covers congestion rents in the global system. And settlement properties are also presented. Finally, numerical results and concluding remarks are shown in Section VI and Section VII.

## II. Interchange Scheduling Schemes

Since loop flows are closely related to interchange scheduling mechanisms, we first review state-of-the-art market mechanisms CTS and GCTS in this section.

### A. Coordinated Transaction Scheduling

Between the borders of many ISOs in the U.S., CTS gets implemented. Market participants bid to buy power at the proxy bus in one area and sell the same amount at the proxy bus in the other area. CTS schedules the interchange level between two neighboring areas with the following steps:

i) Compare LMPs of proxy buses in the two areas when the interchange level is zero. The direction of inter-regional transactions will be from the lower-LMP area to the higher-LMP area.

ii) Each area generates its supply/demand curve ($ISO_1$ as the black curve and $ISO_2$ as the blue curve in Fig. 2) as its proxy-bus LMP under different interchange levels.

iii) The coordinator, which is one of the two ISOs, collects these supply and demand curves, as well as interface bids from market participants. A modified demand curve (green curve in Fig. 2) is obtained by subtracting the aggregated interface bid curve (red curve) from the original demand curve.

iv) The interchange will be set as the intersection of the modified demand curve and the supply curve, if it does not exceed the interchange capacity limit, as in Fig. 2(a). Otherwise, the interchange level will be set as its capacity limit, as in Fig. 2(b).

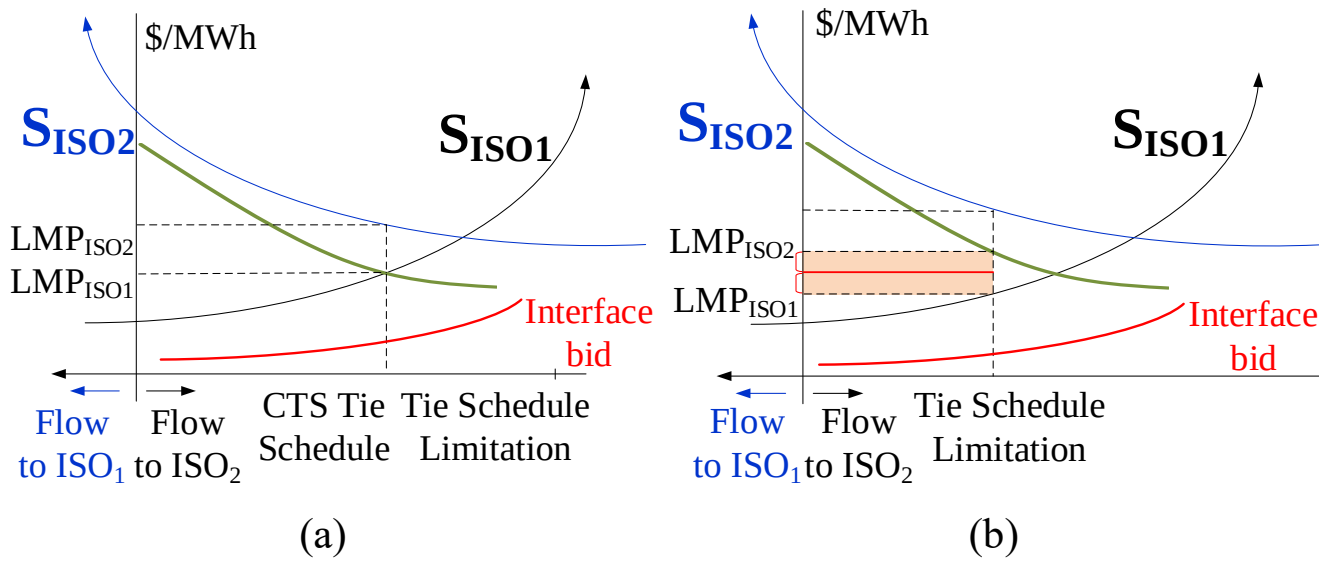

Fig. 2. CTS in the interconnected power system (a) no congestion (b) tie-line congestion

The core design of CTS is to incorporate external market participants to buy and sell at proxy buses of neighboring areas (See Fig. 3). In essence, it is market participants who decide the interchange level and facilitate inter-regional transaction scheduling. Hence, the neutrality of ISOs will be preserved. However, CTS may suffer from the modeling errors from proxy buses, which may lead to inefficient or infeasible tie-line schedules and exacerbate the loop flow problem [23].

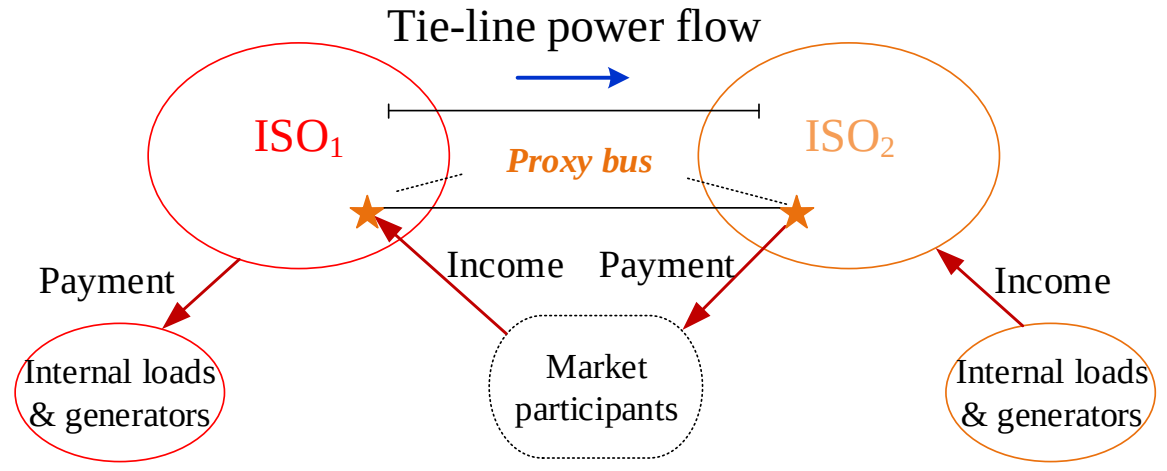

Fig. 3. The financial trading procedure of CTS

### B. Generalized Coordinated Transaction Scheduling

In this paper, we employ the GCTS scheme in [22] as the basic interchange scheduling tool. GCTS preserves the key design of CTS. However, it eliminates proxy buses and allows market participants to bid on any pair of boundary buses. Namely, each interface bid $\Re_b$ is a trio consisting of

$$\Re_b \triangleq \{<B_{pm}, B_{qn}>, \Delta\pi_b, \bar{s}_b\}, \tag{1}$$

where $<B_{pm}, B_{qn}>$ represents that interface bidder $b$ is to buy from bus $p$ of area $m$ and sell to bus $q$ of area $n$, constant $\Delta\pi_b$ is the anticipated price difference between nodes $p$ and $q$, and $\bar{s}_b$ is maximum power quantity.

Based on bids submitted by market participants in the form of (1), the market-clearing model of GCTS is

$$\min_{\boldsymbol{s}, \boldsymbol{g}_i} \quad C = \sum_{i=1}^{N} \boldsymbol{c}_i^T \boldsymbol{g}_i + \Delta\boldsymbol{\pi}^T \boldsymbol{s} \tag{2}$$

$$\boldsymbol{1}^T \boldsymbol{g} = \boldsymbol{1}^T \boldsymbol{d} \qquad \lambda, \tag{3}$$

$$\begin{bmatrix} \boldsymbol{S}_{i,i} \\ \boldsymbol{S}_{t,i} \end{bmatrix} \times (\boldsymbol{g}_i - \boldsymbol{d}_i) + \begin{bmatrix} \boldsymbol{S}_{i,-Bi} \\ \boldsymbol{S}_{t,-Bi} \end{bmatrix} \boldsymbol{M}_{-i} \boldsymbol{s} \le \begin{bmatrix} \boldsymbol{f}_i \\ \boldsymbol{f}_t \end{bmatrix} \qquad \boldsymbol{\mu}_i, \boldsymbol{\mu}_t, \tag{4}$$

$$\underline{\boldsymbol{g}}_i \le \boldsymbol{g}_i \le \bar{\boldsymbol{g}}_i \qquad \underline{\boldsymbol{\eta}}_i, \bar{\boldsymbol{\eta}}_i, \tag{5}$$

$$\boldsymbol{0} \le \boldsymbol{s} \le \bar{\boldsymbol{s}} \qquad \underline{\boldsymbol{\gamma}}, \bar{\boldsymbol{\gamma}}, \tag{6}$$

$$-\boldsymbol{B}_{\bar{i}i} \boldsymbol{B}_{ii}^{-1} (\boldsymbol{g}_i - \boldsymbol{d}_i) = \boldsymbol{M}_i \boldsymbol{s} \qquad \boldsymbol{\rho}_i, \tag{7}$$

where decision variables are generators output $\boldsymbol{g}$ and cleared quantities of interface bids $\boldsymbol{s}$. Vectors $\boldsymbol{c}_i$, $\boldsymbol{d}_i$, $\boldsymbol{f}_i$, $\underline{\boldsymbol{g}}_i$, $\bar{\boldsymbol{g}}_i$ represent, respectively, prices of the generators, load powers, capacity limits of transmission lines, limits of generators output of area $i$. Vector $\boldsymbol{f}_t$ represents the capacity limits of tie-lines. Vector $\bar{\boldsymbol{s}}$ represents the capacity limits of interface bids. Vectors $\boldsymbol{B}_{\bar{i}i}$ and $\boldsymbol{B}_{ii}$ represent mutual-susceptance matrix and self-susceptance matrix of area $i$, respectively. Vector $\boldsymbol{S}_{i,i}$ represents the shift factor between the internal lines and nodes of area $i$. Vector $\boldsymbol{S}_{i,-Bi}$ represents the shift factor between the internal lines and boundary nodes of all areas except area $i$. Vector $\boldsymbol{S}_{t,i}$ represents the shift factor between the tie-lines and internal nodes of area $i$. Vector $\boldsymbol{S}_{t,-Bi}$ represents the shift factor between the tie-lines and boundary nodes of all areas except area $i$. Matrix $\boldsymbol{M}$ is the incidence matrix of interface bids, whose $(i, j)$ entry is equal to one if interface bid buys power at boundary bus $i$, minus one if it sells power at bus $j$, and zero otherwise. Multipliers of all constraints are also listed on the right column of the model.

In the original design of GCTS (as well as CTS), interface bids are cleared in a look-ahead stage but settled at real-time prices. In this paper, for simplicity, we adopt a single-time-scale model by assuming perfect predictions. Consequently, LMPs of the buses in the area $i$ are calculated as

$$\nabla_{\boldsymbol{d}_i} C^* = \lambda \times \boldsymbol{1} - \begin{bmatrix} \boldsymbol{S}_{i,i} \\ \boldsymbol{S}_{t,i} \end{bmatrix}^T \begin{bmatrix} \boldsymbol{\mu}_i \\ \boldsymbol{\mu}_t \end{bmatrix} + (\boldsymbol{B}_{\bar{i}i} \boldsymbol{B}_{ii}^{-1})^T \boldsymbol{\rho}_i \,. \tag{8}$$

The sensitivity of the global optimal cost with respect to the interface bid quantities cleared on the boundary of area $i$ is calculated as

$$\nabla_{\boldsymbol{M}_i \boldsymbol{s}} C^* = \lambda \times \boldsymbol{1} - \begin{bmatrix} \boldsymbol{S}_{i,Bi} \\ \boldsymbol{S}_{t,Bi} \end{bmatrix}^T \begin{bmatrix} \boldsymbol{\mu}_i \\ \boldsymbol{\mu}_t \end{bmatrix} + \boldsymbol{\rho}_i \,, \tag{9}$$

which is equal to the LMP of boundary buses in area $i$. Vector $\boldsymbol{S}_{i,Bi}$ represents the shift factor between the internal lines and boundary nodes of area $i$. Vector $\boldsymbol{S}_{t,Bi}$ represents the shift factor between the tie-lines and boundary nodes of area $i$.

It has been proven in [22] that GCTS can asymptotically converge to the JED solution with sufficiently many interface bidders and if their competitions drive their bidding prices $\Delta\boldsymbol{\pi}$ to zero.

## III. Framework

Loop flow occurs as power flow is strictly governed by physical laws, such as Kirchoff's laws and Ohm's law, and may deviate from the direct path of financial transactions between market participants.

Take the three-area system in Fig. 4 as a more general example. Physically, internal transactions of the area $i$ may also wheel through other areas, leading to loop flow effects on internal lines in areas $j$ and $m$ as well as tie-lines between areas $i$-$j$, $i$-$m$, and $j$-$m$. Similarly, inter-regional transactions between areas $i$-$j$ would lead to loop flow effects on internal lines in areas $i$, $j$, and $m$ as well as tie-lines between areas $i$-$m$ and $j$-$m$. To distribute congestion rents among different areas properly, these components need to be quantified explicitly and clearly.

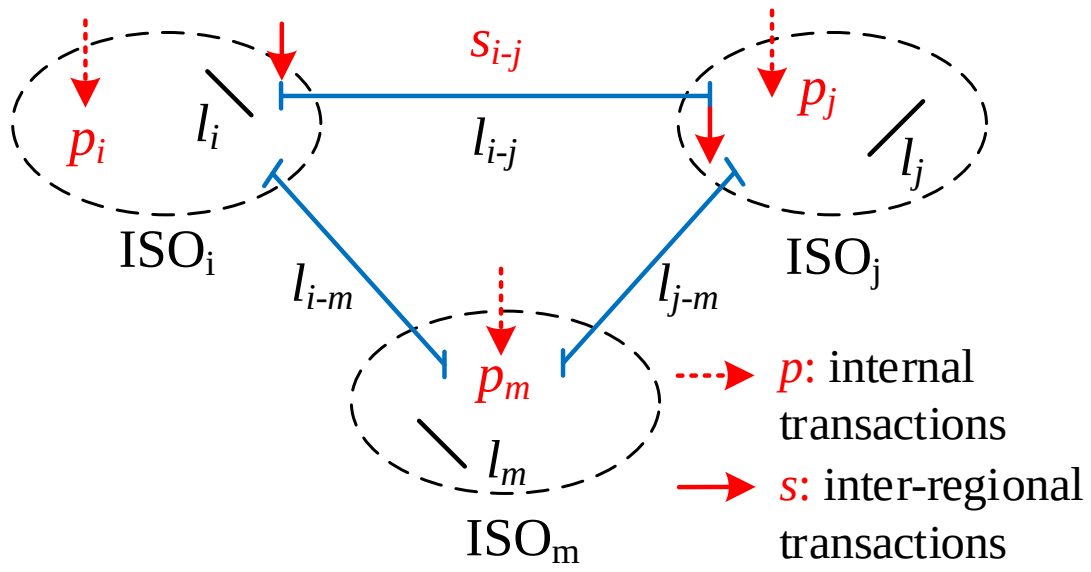

Fig. 4. An illustrative example of loop flow effects

We next describe the framework of jointly covering congestion rents of both internal lines and tie-lines. We assume that cleared quantities of power generations and interface bids are given as inputs. The proposed method has the following steps:

*Step 1*: Recover LMPs by (8) and (9) in a distributed manner.

*Step 2*: Each area settles internal and interface bids, leading to its merchandise surplus.

*Step 3*: Distribute the congestion rent among areas and interface bids according to their contributions.

The second step above is a standard process. Next, we will further explain details of steps 1 and 3, respectively.

## IV. LMP Recovery in a Distributed Manner

Multiple distributed optimization algorithms can be used to obtain the optimal primal solution to (2) ~ (7) in a distributed manner. However, for many of them, each ISO may not be able to obtain optimal dual variables automatically. In this section, we discuss how each area can recover LMPs in its territory that are the same as the global GCTS model with local and boundary equivalent information only. We first adopt linear cost functions for generators, and the case of quadratic cost functions will be discussed separately in subsection *C*.

Namely, assumptions are made as follows:

i) We assume there is no degeneracy in the GCTS model. Degeneracy in each area's internal dispatch problem, however, is still possible under certain conditions. When degeneracies in global problems happen, the method in [24], [25] can be applied.

ii) Each ISO has access to its internal network and equivalent network from all other areas. For example, $\mathrm{ISO_i}$ considers a network as Fig. 5.

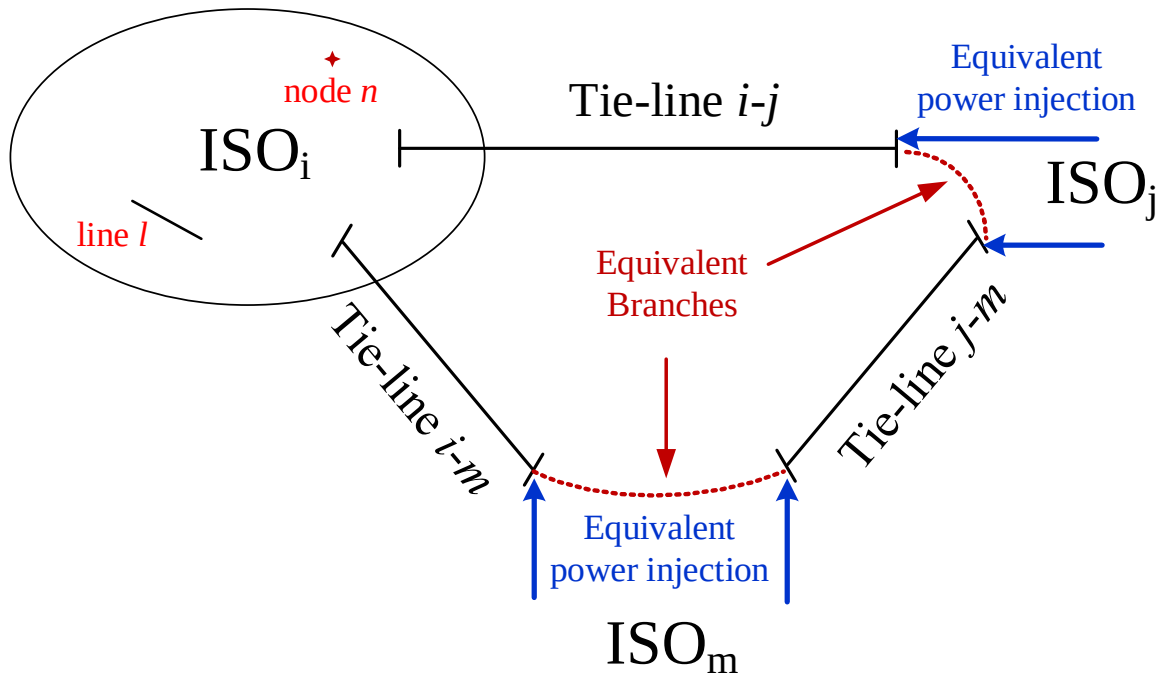

Fig. 5. Network information owned by $\mathrm{ISO_i}$ after doing network equivalence

For simplicity, we start with a simple case without any line congestion to illustrate the idea of distributed LMP recovery.

## A. Illustrative Example

Suppose there is no congestion and no artificial price gap over tie-lines between different areas caused by interchange scheduling. In that case, only one marginal unit decides the same LMP of the entire three-area system. Without loss of generality, we assume the marginal unit is in area *i*. From $\mathrm{ISO_i}$'s point of view, it knows the systemwide LMP which is equal to the bidding price of the marginal unit as shown in Fig. 6 (a). For ISO *j* or *m*, however, the LMP is not uniquely determined since all generators are reaching either their upper or lower limits. Therefore, in this case, $\mathrm{ISO_i}$ needs to share the price of its marginal unit to other areas so that they can perceive the systemwide LMP.

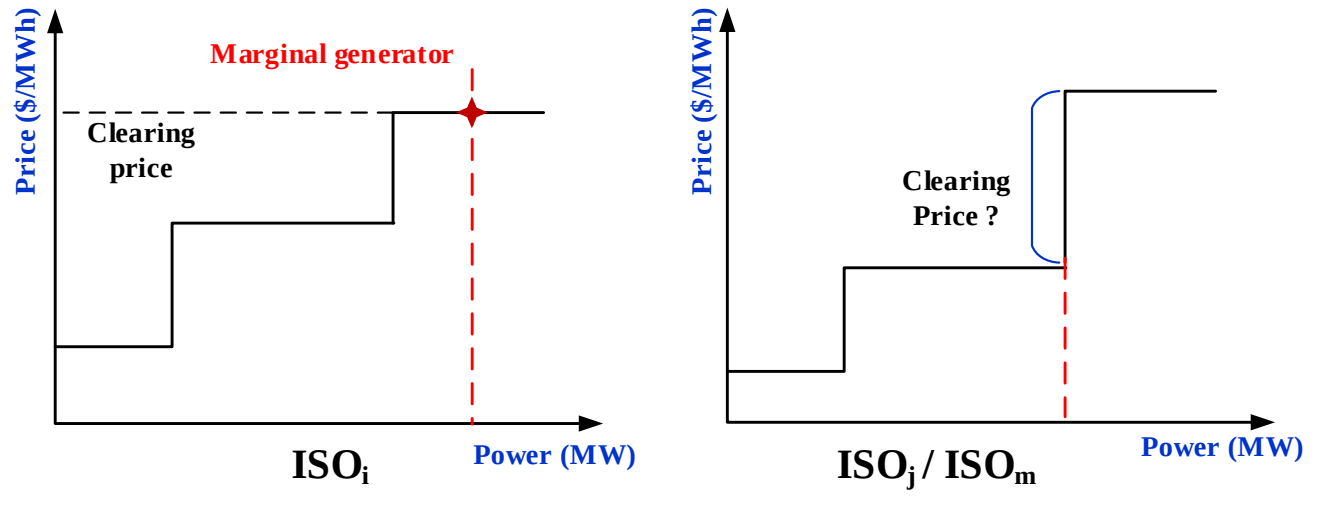

(a) (b)

Fig. 6. Marginal pricing when no degeneracy in the global system (a) $\mathrm{ISO_i}$ (b) $\mathrm{ISO_j}$ and $\mathrm{ISO_m}$

## B. General LMP Recovery Method

For more general cases with possible congestions, we need to calculate the global shift factor matrix in a distributed manner before recovering LMPs. Without loss of generality, we assume that the global phase angle reference bus is selected on the boundary. Entries in the shift factor matrix are classified into two types: intra-regional entries (e.g., transmission line *l* is an internal line in area *i* or a tie-line, and node *n* is in area *i*) and inter-regional entries (e.g., transmission line *l* is in area *i* and node *n* is in area *j*). Each ISO can directly calculate intra-regional entries based on the DC optimal power flow (DCOPF) model with equivalent networks from other areas as in Fig. 5. For the inter-regional entry between line *l* in area *i* and node *n* in area *j* (See Fig. 7), it will be calculated with the following steps:

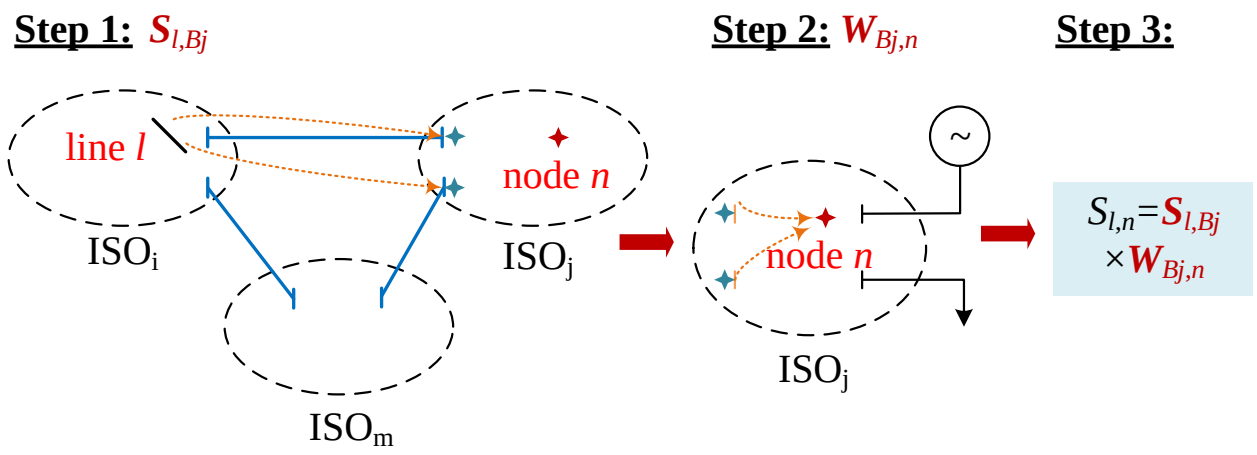

Fig. 7. Calculate inter-regional shift factor entry $S_{l,n}$ in a distributed manner

*Step 1*: Compute the vector of intra-area shift factor entries $\boldsymbol{S}_{l,Bj}$ between line *l* in area *i* and boundary nodes set of area *j*, denoted by *Bj*.

*Step 2*: Compute the coefficient matrix $\boldsymbol{W}_{Bj}=-\boldsymbol{B}_{\bar{j}j}\boldsymbol{B}_{jj}^{-1}$ between boundary nodes and internal nodes of area *j*. The matrix $\boldsymbol{W}_{Bj}$ maps the relationship between boundary equivalent power injections and internal power injections of area *j*.

*Step 3*: Compute inter-regional shift factor entry $S_{l,n}$ by

$$S_{l,n} = \boldsymbol{S}_{l,Bj} \times \boldsymbol{W}_{Bj,n} , \tag{10}$$

where $\boldsymbol{W}_{Bj,n}$ is the column vector corresponding to boundary nodes of area *j* and node *n*.

According to the three steps above, inter-regional shift factor entries will be obtained based on the equivalent network. It should be clarified that we do not need to compute and store all inter-regional shift factor entries. Whenever this entry is needed in the following LMP recovery, we compute it in an on-demand manner.

Next, we explain the general LMP recovery method. According to (8) and (9), LMPs and interface bid prices can be calculated as long as the following multipliers in the GCTS model are recovered:

i) multiplier $\lambda$ associated with constraint (3).

ii) multipliers $\boldsymbol{\mu}_i, \boldsymbol{\mu}_t$ associated with constraint (4).

iii) multiplier $\boldsymbol{\rho}_i$ associated with constraint (7).

We recover these multipliers by looking at marginal generators or marginal interface bids. Marginal prices in (8) and (9) are equal to prices of the marginal generators or marginal interface bid offers, denoted by $\boldsymbol{c}_m$ and $\Delta\boldsymbol{\pi}_m$ respectively, which are known to local operators. Thus we can write a group of linear equations as

$$\frac{\partial L}{\partial \boldsymbol{g}_m} = \boldsymbol{0} \Rightarrow \nabla_{g_m} C^* = \boldsymbol{c}_m , \tag{11}$$

$$\frac{\partial L}{\partial \boldsymbol{s}_m} = \boldsymbol{0} \Rightarrow \nabla_{s_m} C^* = \Delta\boldsymbol{\pi}_m , \tag{12}$$

where *L* is the Lagrangian function and the subscript *m* represents the set of marginal generators or interface bids. Under the non-degeneracy assumption, the number of linearly independent equations in (11) and (12) is equal to the number of variables $\lambda, \boldsymbol{\mu}_i, \boldsymbol{\mu}_t$ and $\boldsymbol{\rho}_i$. Thus linear equations (11) and (12) have a unique solution, which can be used to further recover marginal prices of internal and interchange offers as per (8) and (9). Note that linear equations (11) and (12) need to be solved in a distributed manner, many standard distributed solution algorithms are applicable, such as [26] and [27]. Thereby, we can obtain the solution to multipliers $\lambda$, $\boldsymbol{\mu}_i$, $\boldsymbol{\mu}_t$, $\boldsymbol{\rho}_i$ and each ISO will recover its systemwide LMPs through (8) and (9) based on those multipliers.

### C. *On Cases with Quadratic Cost Functions*

In subsections *A* and *B*, we adopt linear cost functions for generators in the LMP recovery method. Now, we discuss the case of quadratic cost function. The cost function is shown as

$$C = \boldsymbol{g}_i^T \boldsymbol{Q}_i \boldsymbol{g}_i + \boldsymbol{b}_i^T \boldsymbol{g}_i \, , \tag{13}$$

where $\boldsymbol{Q}_i$, $\boldsymbol{b}_i$ represent coefficients of the quadratic cost function. Substituting right sides of equation (11) by the first-order derivative of quadratic cost functions, we obtain the equation as

$$\nabla_{\boldsymbol{g}_m} C^* = 2\boldsymbol{Q}_m \boldsymbol{g}_m^* + \boldsymbol{b}_m \, , \tag{14}$$

where $\boldsymbol{g}_m^*$ is the power output of marginal generators obtained from the market-clearing results. Thus the right side of (14) is also a constant vector. We can still recover LMPs with (12) and (14) when adopting quadratic cost functions.

### D. *LMP Recovery Properties*

The relationship among LMP recovery results, GCTS results, and JED results is analyzed in this subsection. Solving a centralized GCTS model to obtain all LMPs directly requires the information shared among all ISOs. Although distributed algorithms will solve this privacy problem, they do not reveal LMPs directly. Note that the LMP recovery method will obtain the same results as the centralized GCTS based on the generator outputs. Only marginal units and the equivalent boundary system need to be shared. We prove that the proposed LMP recovery results are equal to the GCTS results. The proof is further given in Appendix *A*.

*Theorem 1*: Under the assumption of no degeneracy in the GCTS model (2) ~ (7), the LMP recovered in a distributed manner obtained by (11) and (12) for cases with linear cost functions are equal to those form (8) and (9) calculated with the global GCTS model.

It is noted that even if degeneracy happens, the LMP we recovered still corresponds to their original results. Similarly, with quadratic cost functions, LMPs recovered by (12) and (14) are the same as their counterparts in the global GCTS model. We drop the proof for brevity.

It has already been verified that GCTS results converge to JED results when interface price gaps converge to zero [22]. According to theorem 1, LMP recovery results also converge to JED results when there are sufficiently many interface bids and price gaps over interfaces converge to zero.

## V. Joint Covering Congestion Rent among Each Area and Interface Bids

After recovering global LMPs, each ISO settles internal generators, internal loads, and interface bids accordingly. In this section, we show how system operators and interface bids should jointly cover congestion rents of internal and tie-lines.

### A. *Jointly Covering Congestion Rent*

Physically, tie-line congestions are caused by internal power injections in different areas. In GCTS, however, the external impact of internal transactions is fully captured by interface bids via (7). Consequently, internal transactions in one area will not contribute to tie-line or external line congestions. Thus congestion rents of tie-lines will be covered only by interface bids, whereas congestion rents of internal lines should be covered by internal transactions in the same area and interface bids together. Without loss of generality, the aforementioned three components are denoted by $\psi_{i\text{-}j,m\text{-}n}$, $\psi_{i,i}$, and $\psi_{i,m\text{-}n}$, respectively, where the first subscripts $i$ or $i$-$j$ represents congestion rent of internal lines in area $i$ or that of tie-lines between areas $i$ and $j$, and the second subscripts $i$ or $m$-$n$ shows this is the portion of congestion rent covered by the merchandise surplus of the system operator in area $i$ or an interface bid between areas $m$ and $n$. Note that $i$-$j$ may or may not be equal to $m$-$n$. These three components of congestion rents can be calculated based on shift factors and shadow prices of line congestions [28]:

i) Contribution of an interface bid $s_{m\text{-}n}$ between areas $m$ and $n$ to tie-lines $i$-$j$ is calculated as

$$\psi_{i-j,m-n} = \boldsymbol{\mu}_{i-j}^T \times (\boldsymbol{S}_{i-j,Bm} - \boldsymbol{S}_{i-j,Bn}) s_{m-n} \, , \tag{15}$$

where $\boldsymbol{\mu}_{i\text{-}j}$ is the shadow price vector corresponding to the line capacity constraint of tie-lines $i$-$j$. Based on the LMP recovery method, we obtain entries of $\boldsymbol{\mu}_{i\text{-}j}$ from (11) and (12). Vector $\boldsymbol{S}_{i\text{-}j,Bm}$ is the shift factor between tie-lines $i$-$j$ and the boundary node of area $m$, calculated based on the global equivalent network. Similarly, other shift factor entries in (15), (16), (17) are obtained.

ii) Contribution of internal transactions in area $i$ to its internal lines is calculated as

$$\psi_{i,i} = \boldsymbol{\mu}_i^T \times \boldsymbol{S}_{i,i}(\boldsymbol{g}_i - \boldsymbol{d}_i) \, , \tag{16}$$

where $\boldsymbol{\mu}_i$ is the shadow price vector corresponding to line capacity constraints in area $i$.

iii) Contribution of an interface bid $s_{m\text{-}n}$ between areas $m$ and $n$ to internal lines in area $i$ is calculated as

$$\psi_{i,m-n} = \boldsymbol{\mu}_i^T \times (\boldsymbol{S}_{i,Bm} - \boldsymbol{S}_{i,Bn}) s_{m-n} \, . \tag{17}$$

Take the three-area power system in Fig. 4 as an illustrative example, and its settlement process is in Fig. 8.

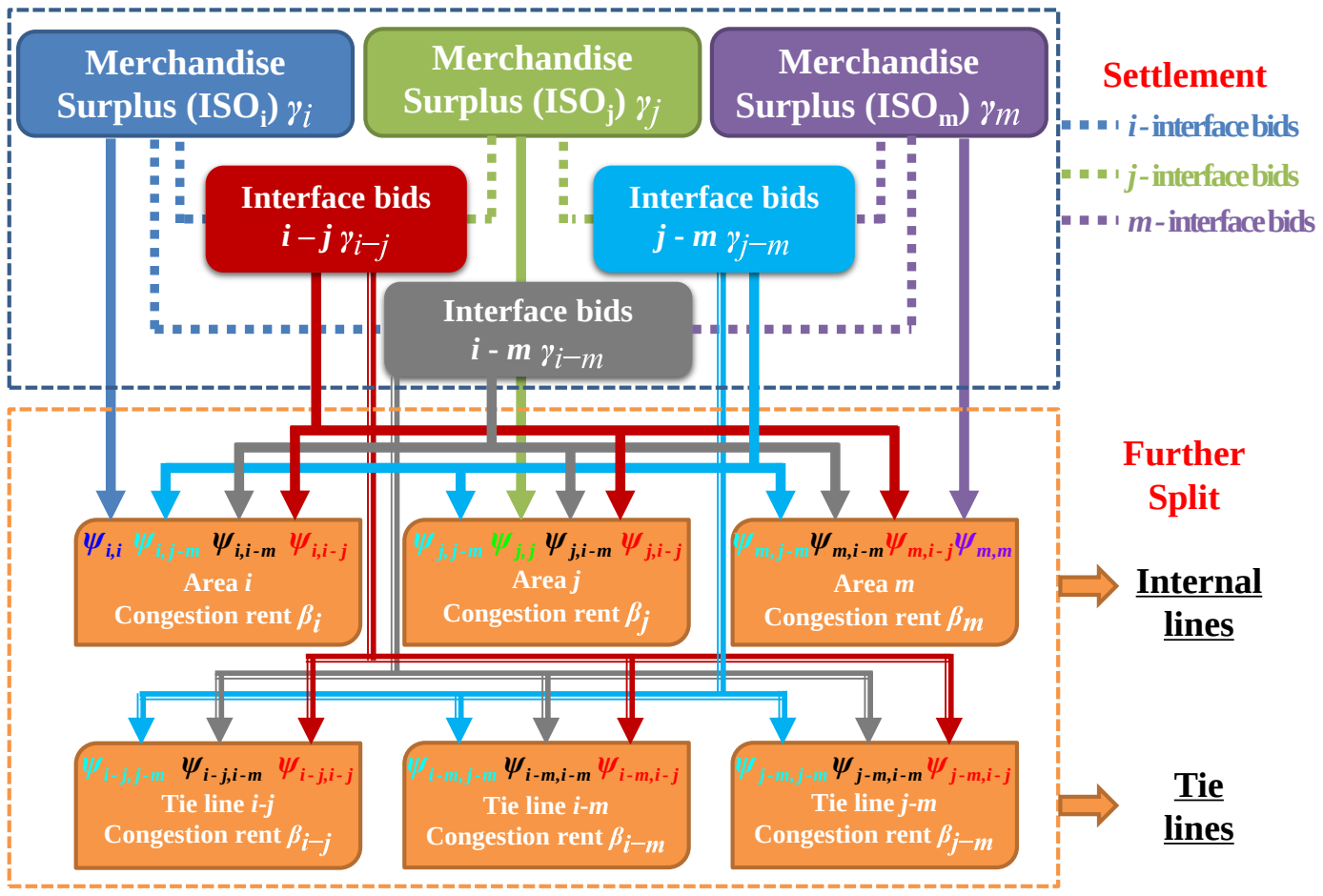

Fig. 8. An illustrative example of jointly covering congestion rents in the three-area power system

In Fig. 8, entries $\beta_i$ represents the congestion rent of internal lines in area $i$. Entries $\beta_{i\text{-}j}$ represents the congestion rent of tie-lines $i$-$j$. Entries $\gamma_i$ represents the congestion rent covered by area $i$. Entries $\gamma_{i\text{-}j}$ represents the congestion rent covered by the interface bid between areas $i$ and $j$. As shown in Fig. 8, there are

$$\beta_i = \psi_{i,i} + \mathbf{1}^T \boldsymbol{\psi}_{i,i-j} + \mathbf{1}^T \boldsymbol{\psi}_{i,i-m} + \mathbf{1}^T \boldsymbol{\psi}_{i,j-m}\,, \tag{18}$$

$$\beta_{i-j} = \mathbf{1}^T \boldsymbol{\psi}_{i-j,i-j} + \mathbf{1}^T \boldsymbol{\psi}_{i-j,i-m} + \mathbf{1}^T \boldsymbol{\psi}_{i-j,j-m}\,, \tag{19}$$

$$\gamma_i = \psi_{i,i}\,, \tag{20}$$

$$\begin{aligned}\gamma_{i-j} &= \psi_{i,i-j} + \psi_{j,i-j} + \psi_{m,i-j} \\ &\quad + \psi_{i-j,i-j} + \psi_{i-m,i-j} + \psi_{j-m,i-j}\end{aligned}\,, \tag{21}$$

where vector $\boldsymbol{\psi}_{i\text{-}j,i\text{-}j}$ includes contributions of all interface bids $i$-$j$ to tie-lines $i$-$j$, and other bold variables in (18), (19) are defined similarly. Entries $\beta_j$, $\beta_m$, $\beta_{i\text{-}m}$, $\beta_{j\text{-}m}$, $\gamma_j$, $\gamma_m$, $\gamma_{i\text{-}m}$, $\gamma_{j\text{-}m}$ in Fig. 8 can be similarly obtained.

## B. Properties

We next establish properties of the settlement process in Fig. 8 as theorems 2, 3, 4. The proofs are further given in Appendix *B*, *C*, *D*, respectively.

*Theorem 2*: Under the assumption of no degeneracy in GCTS, congestion rent of tie-lines and that of internal lines in each area are equal to the amount calculated by the joint economic dispatch problem:

$$\beta_{i-j} = \boldsymbol{\mu}_{i-j}^T \boldsymbol{f}_{i-j},\quad \beta_i = \boldsymbol{\mu}_i^T \boldsymbol{f}_i\,, \tag{22}$$

where $\boldsymbol{f}_{i\text{-}j}$, $\boldsymbol{f}_i$ represent the power flows of tie-lines $i$-$j$ and internal lines in area $i$, respectively. Even if degeneracy occurs, as long as equations corresponding to multipliers are available through [24] and [25], we can recover LMPs to settle each area and interface bids. Then this theorem still holds.

*Theorem 3*: Under the assumption of no degeneracy in GCTS, the congestion rent afforded by ISO$_i$ is equal to its merchandise surplus. And for an interface bid buying from area $j$ and selling to the area $i$, the congestion rent afforded by it is equal to its profit:

$$\gamma_i = MS_i\,, \tag{23}$$

$$\gamma_{i-j} = R_{i-j} - \Delta\pi_{m,i-j} s_{i-j}\,, \tag{24}$$

where entry $MS_i$ represents the merchandise surplus of area $i$, entry $R_{i\text{-}j}$ represents the arbitrage revenue of a given interface bid buying from area $j$ and selling to the area $i$, and entry $\Delta\pi_{m,\,i\text{-}j}$ represents the anticipated price difference bid by the marginal interface bid between areas $i$-$j$. Similar to theorem 2, this theorem still holds under the degeneracy case.

It should be noticed that, in general, shift factors will change with the selection of phase angle reference bus, whereas shadow prices $\boldsymbol{\mu}$ and congestion rents do not. In the multi-area problem, the proposed scheme should not be affected by that either. In other words, all congestion rent components in (15) ~ (17) should be independent to the selection of phase angle reference bus. Next, we will elaborate that such a property holds in the proposed mechanism.

When we calculate $\psi_{i\text{-}j,m\text{-}n}$, the difference term $\boldsymbol{S}_{i\text{-}j,Bm}$-$\boldsymbol{S}_{i\text{-}j,Bn}$ in (15) represents the shift factor between tie-lines $i$-$j$ and the interface bid $s_{m\text{-}n}$. For interface bid $s_{m\text{-}n}$, when every *MW* injects at one node, a corresponding *MW* withdraws at the other node. The compensation effects from the phase angle reference bus cancel out. Thus the calculation result of entry $\psi_{i\text{-}j,m\text{-}n}$ is independent to the selection of slack bus. And the calculation result of the entry $\psi_{i,m\text{-}n}$ in (17) is similar.

We next illustrate that the calculation result of $\psi_{i,i}$ in (16) is also independent to the selection of phase angle reference bus. In order to make $\boldsymbol{S}_{i,i}$ in (16) independent to that either, interface bids will play the role of slack bus proportionally. Note that we have already assumed that there is no generator or load at the boundary node, meaning that the corresponding power injection is equal to zero. Take area $i$ in Fig. 9 as an example, where areas $j$ and $m$ have been represented by the equivalent network. The interface bid $s_{B1\text{-}B3}$ withdraws power from the boundary node $B3$, namely supplying power for area $j$. This power is bought from the boundary node $B1$ in area $i$, and it is physically provided by internal nodes of area $i$ proportionally. It is similar for the interface bid $s_{B2\text{-}B4}$. We calculate the quantities of interface bids $s_{B1\text{-}B3}$ and $s_{B2\text{-}B4}$ by

$$s_{B1-B3} = -\boldsymbol{B}_{B1,i}\boldsymbol{B}_{ii}^{-1}(\boldsymbol{g}_i - \boldsymbol{d}_i)\,, \tag{25}$$

$$s_{B2-B4} = -\boldsymbol{B}_{B2,i}\boldsymbol{B}_{ii}^{-1}(\boldsymbol{g}_i - \boldsymbol{d}_i)\,, \tag{26}$$

which are consistent with the definition of interface bid in (7). Given the above setting, when the power injection at any node $n$ ($n$=1, 2, 3) in area $i$ changes, the equivalent power injections at nodes $B1$ and $B2$ will change proportionally according to (25), (26). Therefore, the calculation result of the shift factor matrix $\boldsymbol{S}_{i,i}$ does not depend on the artificially selected phase angle reference bus. That is to say, $\psi_{i,i}$ in (16) is independent to the selection of phase angle reference bus.

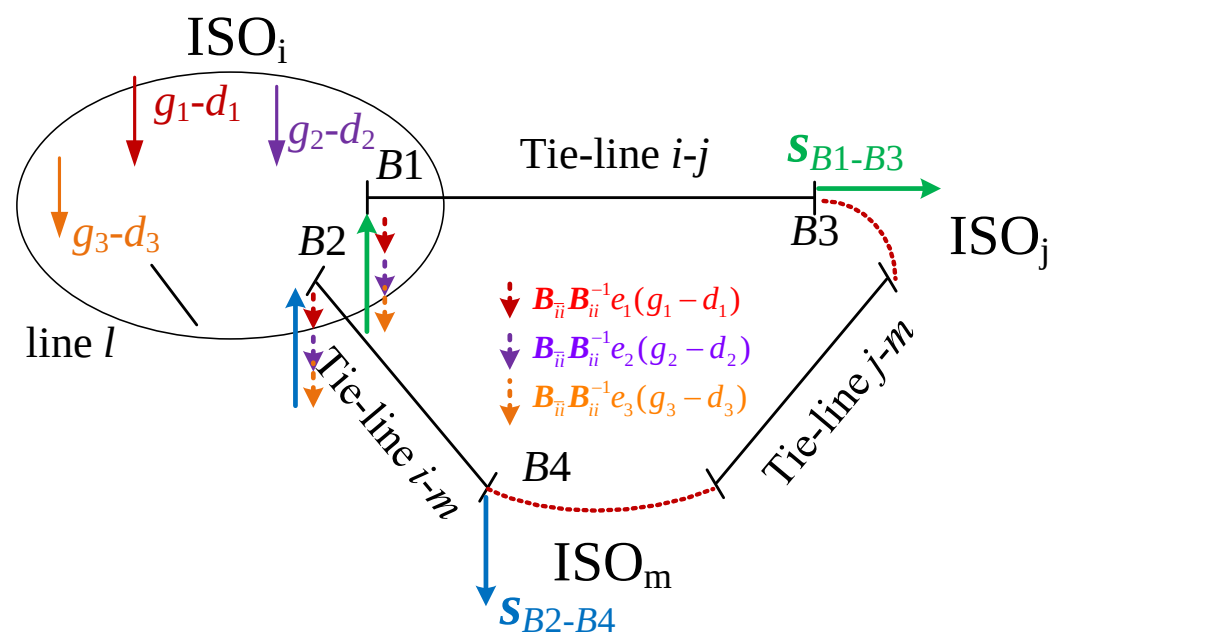

Fig. 9. Example of calculating the contribution of internal transactions of area $i$ to its internal lines

Next, we calculate entries of the shift factor matrix $\boldsymbol{S}_{i,i}$ between internal lines and nodes in area $i$ as follows.

*Step 1*: Compute the shift factor entry $S_{l,n}$ between internal line $l$ and node $n$ of area $i$ with a phase angle reference bus.

*Step 2*: Compute the shift factor $\boldsymbol{S}_{l,Bi}$ between internal line $l$ and boundary nodes of area $i$, denoted by $Bi$.

*Step 3*: Compute the weight coefficient vector $\boldsymbol{B}_{\bar{i}i}\boldsymbol{B}_{ii}^{-1}$ based on network equivalence.

*Step 4*: Compute the shift factor entry of the matrix $\boldsymbol{S}_{i,i}$ by

$$S_{l,n} = (S_{l,n}\mathbf{1} - \boldsymbol{S}_{l,Bi})\boldsymbol{B}_{\bar{i}i}\boldsymbol{B}_{ii}^{-1}\,. \tag{27}$$

Similar to (15) and (17), the difference term $\boldsymbol{S}_{l,n}\mathbf{1}$-$\boldsymbol{S}_{l,Bi}$ in (27) also removes the compensation effects from the phase angle reference bus. Thus, the calculation results of the shift factor $\boldsymbol{S}_{i,i}$ and component $\psi_{i,i}$ are independent to the selection of phase angle reference bus.

Note that in [22], internal congestion rents are solely covered by the local ISO, and interface bids cover tie-line congestion rents. However, in [22] interface bids are settled at a different price from the LMP of boundary buses, which is not easy to understand. In this paper, by defining prices of interface bids with equation (9), which is equal to the LMP at the same boundary bus, and explicitly calculating components of congestion rents, formerly obscure interface bid prices become understandable: We settle interface bids according to the boundary bus LMPs, then subtract the internal and tie-line congestion rents covered by them. In the following theorem, we show that the proposed scheme is equivalent to that in [22].

*Theorem 4:* Under the assumption of no degeneracy in GCTS, we have

$$\nabla_s^T C_i^* - \boldsymbol{\mu}_i^T \boldsymbol{S}_{i,B}\boldsymbol{M} = \nabla_{\boldsymbol{M}_i s}^T C^* \boldsymbol{M}_i \,, \qquad (28)$$

where $\boldsymbol{S}_{i,B}$ represents the shift factor between the internal lines of area $i$ and boundary nodes. We prove this theorem in Appendix *D*. Similar to theorem 2, this theorem still holds when degeneracy occurs.

## VI. Numerical Experiments

The performance of the proposed method is evaluated with the two-area (4 nodes) power system in Fig. 10, the two-area (23 nodes) power system in Fig. 12, and the three-area (28 nodes) power system in Fig. 14, respectively. Simulations are carried out on a PC with an Intel Core i5 processor running at 3.00 GHz with 12 GB of memory.

### *A. Case 1: Two-Area (4 Nodes) Power System*

We first use a two-area (4 nodes) system in Fig. 10 to solve the GCTS model and jointly cover the congestion rent among areas and interface bids. Parameters of generators and loads are listed in Table I. Without loss of generality, node 1 is selected as the common phase angle reference bus of two areas. The capacity of tie-line (1-3) is set as 10 MW, while capacities of other lines are set as infinity. Reactances of lines (1-2), (1-3), (3-4) are all set as 1.0 p. u.. We set a variable reactance $x_{2\text{-}4}$ for line (2-4) to consider the network topology with or without a loop. When $x_{2\text{-}4}$ is set as infinity, there is no loop in the network. The profile of interface bids is shown in Table II, including bidding locations, maximum power quantities, and the anticipated bidding prices.

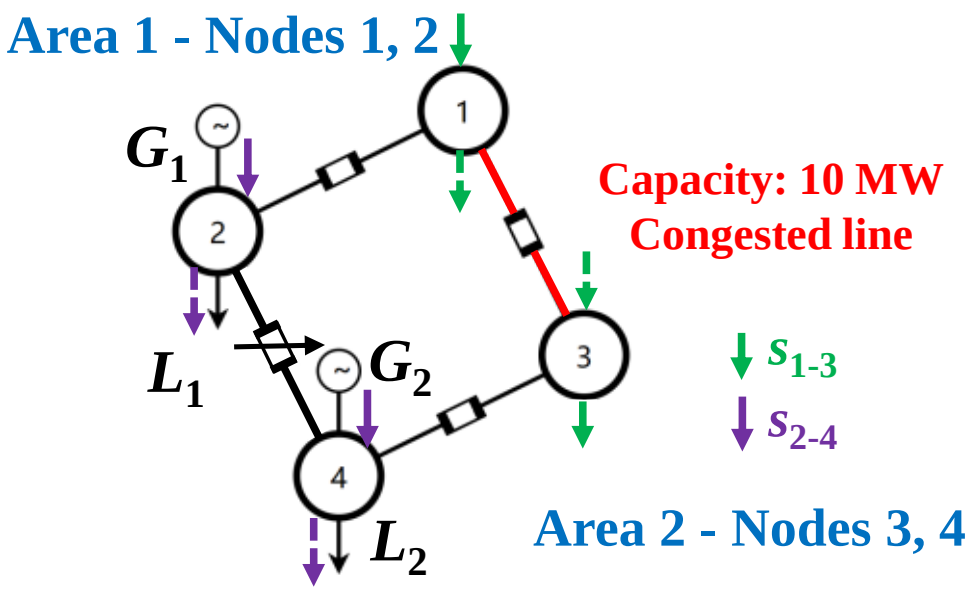

Fig. 10. The two-area power system in case 1

Table I
Parameters of Generators and Loads in Case 1

| $c_1$ (\$/MWh) | $\underline{g}_1$ (MW) | $\overline{g}_1$ (MW) | $c_2$ (\$/MWh) | $\underline{g}_2$ (MW) | $\overline{g}_2$ (MW) | $L_1$ (MW) | $L_2$ (MW) |
|---|---|---|---|---|---|---|---|
| 1.0 | 0.0 | 100.0 | 2.0 | 0.0 | 100.0 | 30.0 | 60.0 |

Table II
Profile of Interface Bids in Case 1

| Indices | Sell to (Area-Bus) | Buy from (Area-Bus) | Price (\$/MWh) | Upper limit of bidding quantity (MW) |
|---|---|---|---|---|
| $s_{1\text{-}3}$ | 2-3 | 1-1 | 0.0 | 200 |
| $s_{2\text{-}4}$ | 2-4 | 1-2 | 0.0 | 200 |

Next, we use the case of setting $x_{2\text{-}4}$ as infinity to explain the settlement process of joint covering congestion rent. When $x_{2\text{-}4}$ is set as 1.0 p. u., the process is similar and omitted here.

First, we solve the GCTS model. Market-clearing results are listed in Table III, including LMPs, generator outputs, and net revenue of each node. Clearing results of interface bids are listed in Table IV, including cleared quantities, tie-line flow, and marginal prices. The shadow price of the capacity constraint corresponding to the congested line (1-3) is 1.0 \$/MWh. Congestion rent of the global system is 10.0 \$/h.

Table III
Market-Clearing Results in Case 1

| Reactance (p. u.) | $x_{2\text{-}4}=\infty$ | | | | $x_{2\text{-}4}=1.0$ | | | |
|---|---|---|---|---|---|---|---|---|
| Node | Bus 1 | Bus 2 | Bus 3 | Bus 4 | Bus 1 | Bus 2 | Bus 3 | Bus 4 |
| LMP (\$/MWh) | 1.0 | 1.0 | 2.0 | 2.0 | 0.0 | 1.0 | 3.0 | 2.0 |
| Generator output (MW) | / | 40.0 | / | 50.0 | / | 70.0 | / | 20.0 |
| Net revenue (\$/h) | 0.0 | -10.0 | 0.0 | 20.0 | 0.0 | -40.0 | 0.0 | 80.0 |

Table IV
Clearing Results of Interface Bids in Case 1

| Reactance (p. u.) | $x_{2\text{-}4}$ | $\infty$ | 1.0 |
|---|---|---|---|
| Cleared quantities of interface bid (MW) | $s_{1\text{-}3}$ | 10.0 | 0.0 |
| | $s_{2\text{-}4}$ | 0.0 | 40.0 |
| Tie-line flow (MW) | Bus 1 to 3 | 10.0 | 10.0 |
| | Bus 2 to 4 | 0.0 | 30.0 |
| Marginal prices (\$/MWh) | Bus 1 | 1.0 | / |
| | Bus 2 | / | 1.0 |
| | Bus 3 | 2.0 | / |
| | Bus 4 | / | 2.0 |

Second, according to the market-clearing results of GCTS, area 1, area 2, and interface bids are settled. Take area 1 as an example. It collects 30 \$/h from internal load $L_1$ and 10 \$/h from interface bid $s_{1\text{-}3}$. And it pays 40 \$/h for the internal generator $G_1$. The merchandise surplus of area 1 is 0 \$/h. The settlement of area 2 is similar. The interface bid $s_{1\text{-}3}$ collects 20 \$/h from area 2 and pays 10 \$/h for area 1. Thus, the profit of $s_{1\text{-}3}$ is 10 \$/h. The global congestion rent equals the sum of merchandise surpluses of areas 1, 2, and the profit of interface bid $s_{1\text{-}3}$.

Third, we will quantify the contribution of interface bids and internal transactions to the line (1-3) and show how to jointly cover congestion rents based on shadow prices. In Fig. 10, the interface bid $s_{1\text{-}3}$ withdraws 10 MW power from node 3, namely supplying 10 MW power for area 2. This power is bought from node 1 in area 1, and it is physically provided by node 2. In area 1, when the power injection at node 2 increases every *MW* power, the equivalent power withdrawal at node 1 will increase a corresponding *MW* power. In area 2, it is similar for nodes 3 and 4. Thus, shift factor entries between tie-line (1-3) and $s_{1\text{-}3}$, $g_1$-$L_1$, $g_2$-$L_2$ are 1, 0, 0, respectively. By multiplying the quantities of $s_{1\text{-}3}$, $g_1$-$L_1$, $g_2$-$L_2$ with shift factor entries, their contributions to tie-line (1-3) are 10 MW, 0 MW, 0 MW. Then, by multiplying the contributions with the shadow price of tie-line (1-3) 1.0 $/MWh, we obtain the congestion rents afforded by interface bid $s_{1\text{-}3}$, area 1, area 2 as 10 $/h, 0 $/h, 0 $/h, respectively.

In this case, results of joint covering congestion rents among area 1, area 2, and interface bid are shown in Fig. 11 and Table V, which gives a clear answer to the example in Fig. 1.

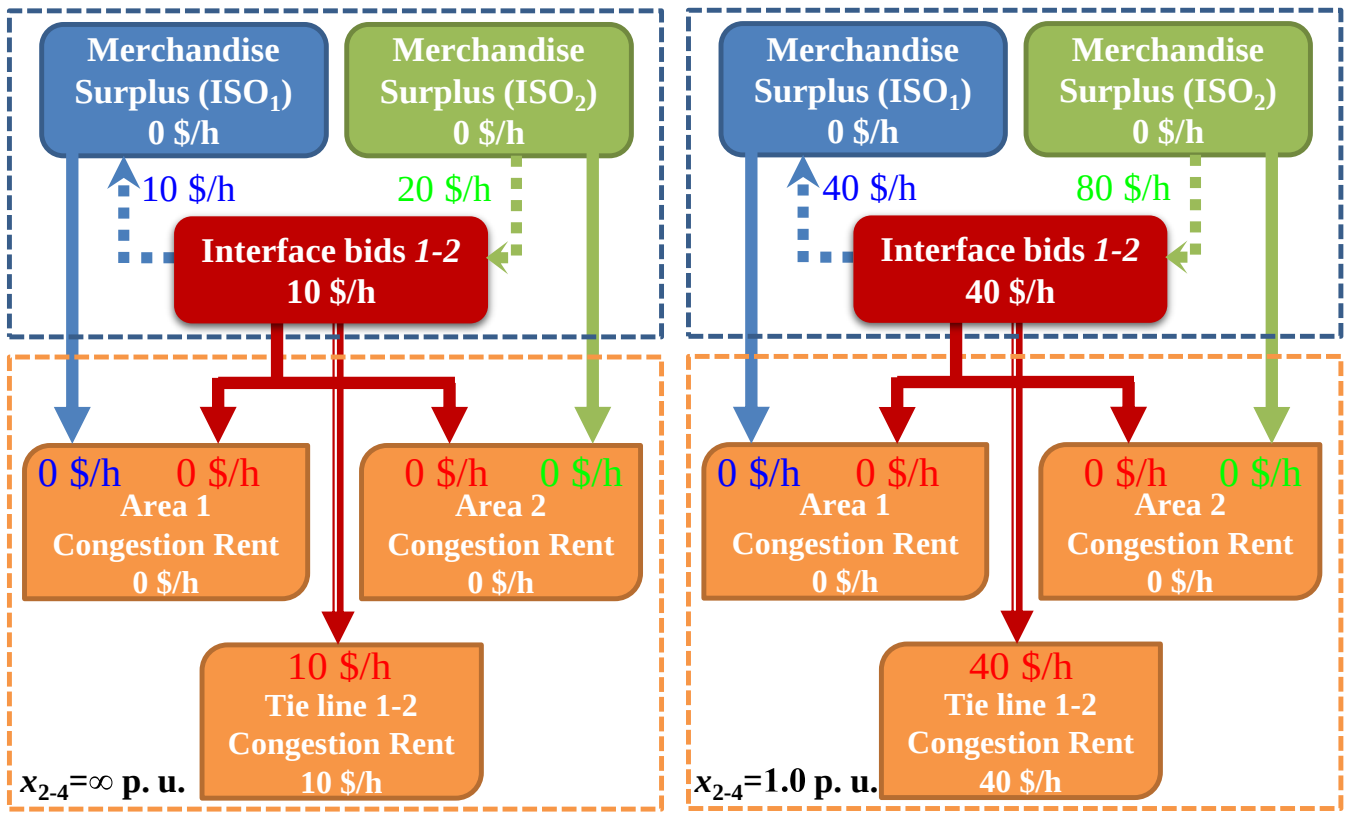

Fig. 11. Results of jointly covering congestion rents in case 1

TABLE V
SETTLEMENT OF AREA 1, AREA 2, AND INTERFACE BIDS IN CASE 1

| Reactance (p. u.) | Settlement process | Area 1 | Area 2 | Interface bid |
|---|---|---|---|---|
| | Collect from internal generators ($/h) | -40.0 | -100.0 | 0.0 |
| | Collect from internal loads ($/h) | 30.0 | 120.0 | 0.0 |
| $x_{2\text{-}4}$=∞ | Collect from interface bids ($/h) | 10.0 | -20.0 | 10.0 |
| | Merchandise surpluses ($/h) | 0.0 | 0.0 | 10.0 |
| | Congestion rent afforded ($/h) | 0.0 | 0.0 | 10.0 |
| | Collect from internal generators ($/h) | -70.0 | -40.0 | 0.0 |
| | Collect from internal loads ($/h) | 30.0 | 120.0 | 0.0 |
| $x_{2\text{-}4}$=1.0 | Collect from interface bids ($/h) | 40.0 | -80.0 | 40.0 |
| | Merchandise surpluses ($/h) | 0.0 | 0.0 | 40.0 |
| | Congestion rent afforded ($/h) | 0.0 | 0.0 | 40.0 |

### *B. Case 2: Two-Area (23 Nodes) Power System*

We next use a power system of 23 nodes (See Fig. 12) to verify the effectiveness of the proposed approaches. The case is composed of IEEE 14, 9-bus system and two extra tie-lines. We provide parameters of generators and tie-lines in Table VI and Table VII. The profile of interface bids is described in Table VIII. We adopt six cases (S0 ~ S5) set as follows.

**S0**: Capacities of transmission lines are set as infinity, and bidding prices of interface bids are 0.0 $/MWh.

**S1**: Capacities of transmission lines are set as infinity, and bidding prices of interface bids are 0.1 $/MWh.

**S2**: Capacity of tie-line (10-20) is set as 5 MW, and capacities of other transmission lines are infinity. The bidding prices of interface bids are 0.0 $/MWh.

**S3**: Capacity of tie-line (10-20) is set as 5 MW, and capacities of other transmission lines are infinity. The bidding prices of interface bids are 0.1 $/MWh.

**S4**: Capacity of the internal line (9-10) is set as 5 MW, and capacities of other transmission lines are infinity. The bidding prices of interface bids are 0.0 $/MWh.

**S5**: Capacity of the internal line (9-10) is set as 5 MW, and capacities of other transmission lines are infinity. The bidding prices of interface bids are 0.1 $/MWh.

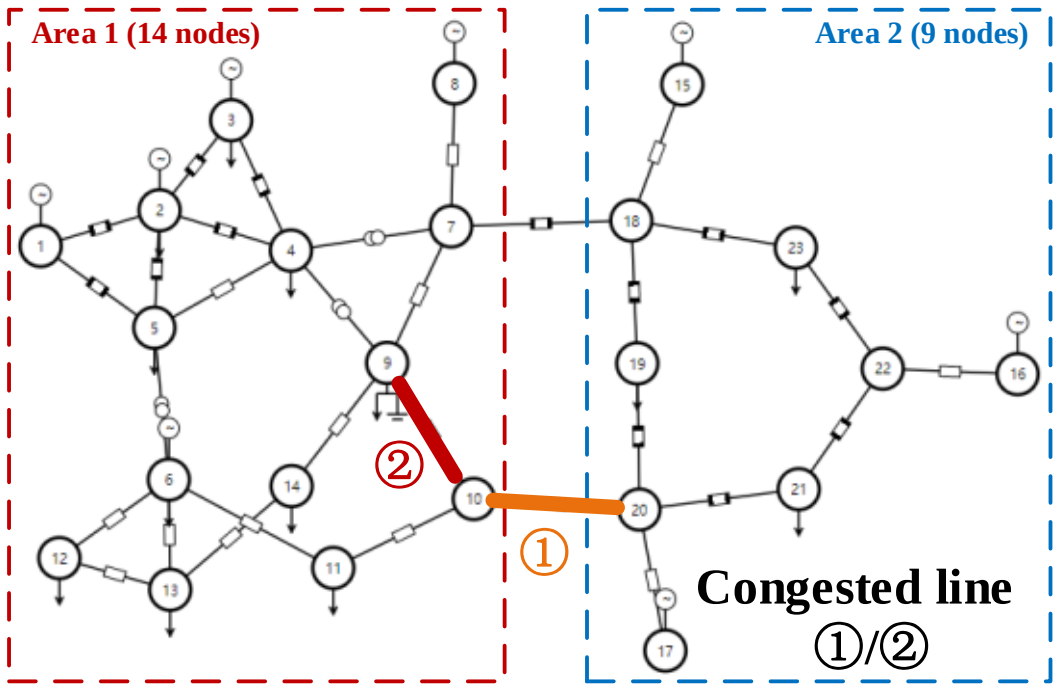

Fig. 12. The two-area power system in case 2

TABLE VI
PARAMETERS OF GENERATORS IN CASE 2

| Number | Location (Area-Bus) | Cost coefficient ($/MWh) | Lower limit (MW) | Upper limit (MW) |
|---|---|---|---|---|
| 1 | 1-1 | 0.5 | 0.0 | 332.4 |
| 2 | 1-2 | 2.0 | 0.0 | 140.0 |
| 3 | 1-3 | 3.0 | 0.0 | 100.0 |
| 4 | 1-6 | 4.0 | 0.0 | 100.0 |
| 5 | 1-8 | 5.0 | 0.0 | 100.0 |
| 6 | 2-15 | 50.0 | 0.0 | 100.0 |
| 7 | 2-16 | 12.0 | 0.0 | 100.0 |
| 8 | 2-17 | 8.0 | 0.0 | 100.0 |

TABLE VII
TIE-LINE PARAMETERS IN CASE 2

| Tie-line Number | Sending node (Area-Bus) | Receiving node (Area-Bus) | Reactance (p. u.) | Capacity (MW) |
|---|---|---|---|---|
| TL1 | 1-7 | 2-18 | 0.161 | 500 |
| TL2 | 1-10 | 2-20 | 0.085 | 500 |

TABLE VIII
PROFILE OF INTERFACE BIDS IN CASE 2

| Indices | Sell to (Area-Bus) | Buy from (Area-Bus) | Upper limit of bidding quantity (MW) |
|---|---|---|---|
| 1 | 2-18 | 1-7 | 200 |
| 2 | 2-18 | 1-10 | 250 |
| 3 | 2-20 | 1-7 | 300 |
| 4 | 2-20 | 1-10 | 350 |
| 5 | 1-7 | 2-18 | 200 |
| 6 | 1-10 | 2-18 | 250 |
| 7 | 1-7 | 2-20 | 300 |
| 8 | 1-10 | 2-20 | 350 |

For this two-area system, network equivalence is applied individually. As a result, area 1 includes 16 nodes (12 internal nodes and 4 boundary nodes) and area 2 contains 11 nodes (7 internal nodes and 4 boundary nodes). Without loss of generality, node 18 on the boundary system is adopted as the common phase angle reference bus.

First, we solve the GCTS model of cases S0 ~ S5 separately. Generator outputs are listed in Table IX. Clearing results of interface bids are shown in Table X, including cleared quantities of interface bids, tie-line flow, marginal prices, and market cost. Here, the market cost includes both generation cost and the cost of clearing interface bids.

TABLE IX
GENERATOR OUTPUTS IN CASE 2

| Indices | Generator outputs S0, S1 (MW) | Generator outputs S2, S3 (MW) | Generator outputs S4, S5 (MW) |
|---|---|---|---|
| 1 | 332.4 | 332.4 | 332.4 |
| 2 | 140.0 | 11.3 | 0.0 |
| 3 | 92.6 | 0.0 | 0.0 |
| 4 | 0.0 | 0.0 | 62.1 |
| 5, 6, 7 | 0.0 | 0.0 | 0.0 |
| 8 | 0.0 | 221.3 | 172.5 |

TABLE X
CLEARING RESULTS OF INTERFACE BIDS IN CASE 2

| Cases | | S0 | S1 | S2 | S3 | S4 | S5 |
|---|---|---|---|---|---|---|---|
| Cleared quantities of interface bids (MW) | 1 | 122.3 | 134.1 | 110.3 | 113.6 | 111.1 | 111.2 |
| | 2 | 110.2 | 98.5 | 0.0 | 0.0 | 0.0 | 0.0 |
| | 3 | 59.8 | 48.0 | 71.6 | 68.5 | 71.0 | 70.9 |
| | 4 | 22.7 | 34.5 | 0.0 | 0.0 | 0.0 | 0.0 |
| | 5 | 0.0 | 0.0 | 0.0 | 0.0 | 0.0 | 0.0 |
| | 6 | 0.0 | 0.0 | 29.3 | 32.6 | 13.0 | 13.1 |
| | 7 | 0.0 | 0.0 | 0.0 | 0.0 | 0.0 | 0.0 |
| | 8 | 0.0 | 0.0 | 59.1 | 55.8 | 26.6 | 26.5 |
| Tie-line flow (MW) | Bus 7 to 18 | 164.1 | 164.1 | 88.7 | 88.7 | 100.8 | 100.8 |
| | Bus 10 to 20 | 150.9 | 150.9 | 5.0 | 5.0 | 41.7 | 41.7 |
| Marginal prices ($/MWh) | Bus 7 | 3.0 | 3.0 | 2.8 | 2.8 | 1.4 | 1.3 |
| | Bus 10 | 3.0 | 3.0 | 0.3 | 0.3 | 9.8 | 9.9 |
| | Bus 18 | 3.0 | 3.1 | 5.4 | 5.6 | 0.8 | 4.8 |
| | Bus 20 | 3.0 | 3.1 | 8.0 | 8.0 | 4.7 | 8.0 |
| Market cost ($/h) | | 1809 | 1841 | 3044 | 3071 | 2872 | 2894 |

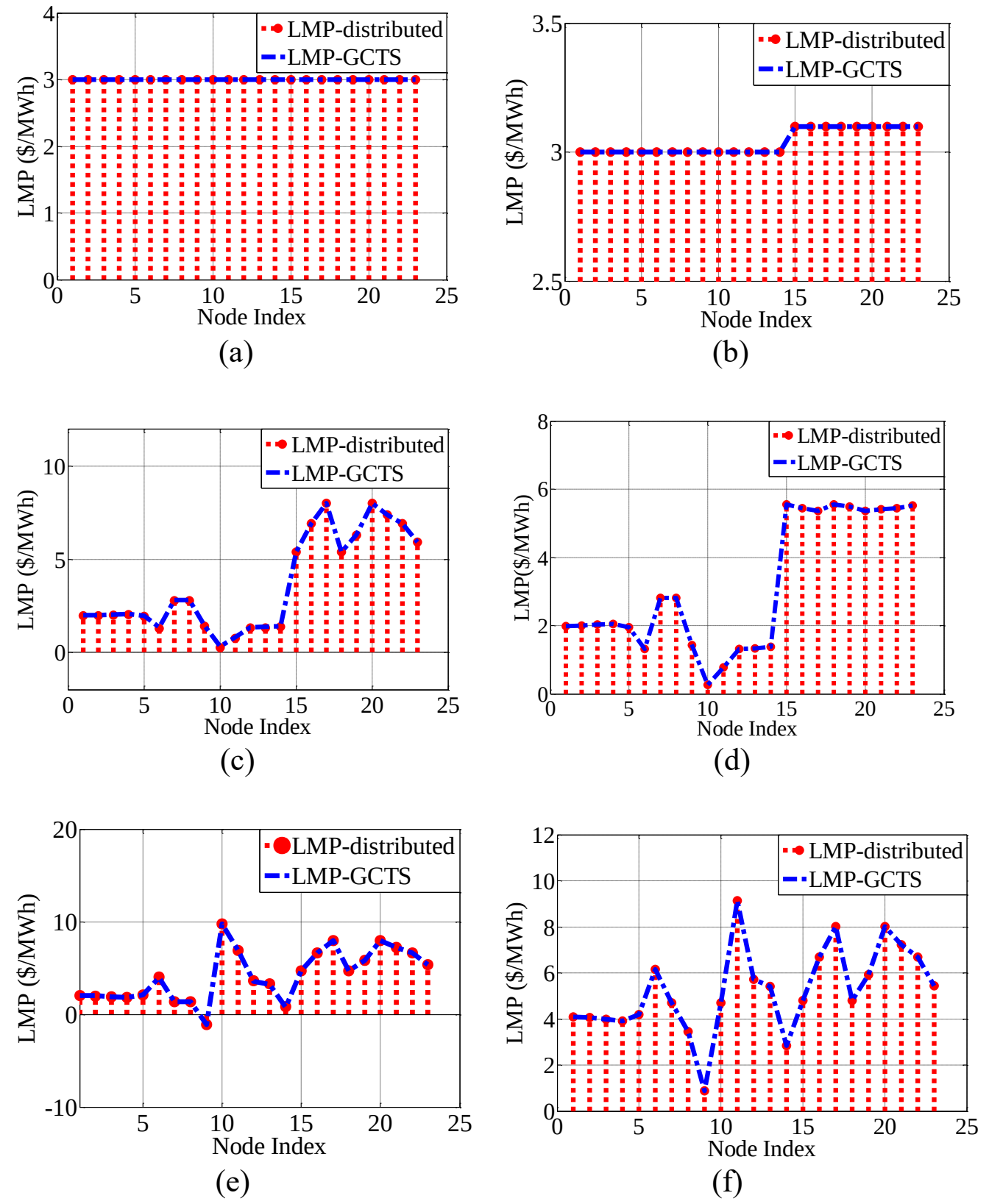

Fig. 13. Comparison between LMP recovery and GCTS when cost functions are linear (a) S0 (b) S1 (c) S2 (d) S3 (e) S4 (f) S5

Second, we will verify the effectiveness of the LMP recovery method. When cost functions are linear, the distributed LMP recovery method reveals the same results as the GCTS solved in a centralized manner, which is shown in Fig. 13. Compared with obtaining LMPs from the centralized GCTS model, the LMP recovery method is based on marginal units and the shift factor entries obtained in a distributed manner so that the privacy of each area is better protected.

Third, after recovering the global LMPs, we further verify the approach of jointly covering global congestion rent. Settlements of each area and interface bids under cases S0 ~ S5 are shown in Table XI. For S0 and S1, there are no line congestions in the global system. Congestion rents covered by area 1, 2, interface bids 1-2 are all 0 $/h. For S2 and S3, congestion rent of tie-line (10-20) would be covered by interface bids 1-2. For S4 and S5, congestion rent of internal line (9-10) in area 1 would be covered by area 1 and interface bids 1-2 proportionally. Under cases S0 ~ S5, the merchandise surplus of area 1 is equal to congestion rent afforded by area 1. Results of area 2 are similar. And the profit of interface bids 1-2 is equal to congestion rent afforded by them.

TABLE XI
SETTLEMENT OF AREA 1, AREA 2, AND INTERFACE BIDS IN CASE 2

| Case ID | Areas/ bids | Collect from internal generators and loads ($/h) | Collect from interface bids ($/h) | Merchandise surplus ($/h) | Cost of clearing interface bids ($/h) | Congestion rent afforded ($/h) |
|---|---|---|---|---|---|---|
| S0 | Area1 | -364.19 | 364.19 | 0.00 | / | 0.00 |
| | Area2 | 364.19 | -364.19 | 0.00 | / | 0.00 |
| | $s_{1\text{-}2}$ | 0.00 | 0.00 | 0.00 | 0.00 | 0.00 |
| S1 | Area1 | -364.19 | 364.19 | 0.00 | / | 0.00 |
| | Area2 | 382.40 | -382.40 | 0.00 | / | 0.00 |
| | $s_{1\text{-}2}$ | 0.00 | 18.21 | 18.21 | 18.21 | 0.00 |
| S2 | Area1 | -230.90 | 230.90 | 0.00 | / | 0.00 |
| | Area2 | 276.31 | -276.31 | 0.00 | / | 0.00 |
| | $s_{1\text{-}2}$ | 0.00 | 45.41 | 45.41 | 0.00 | 45.41 |
| S3 | Area1 | -231.62 | 231.62 | 0.00 | / | 0.00 |
| | Area2 | 304.84 | -304.84 | 0.00 | / | 0.00 |
| | $s_{1\text{-}2}$ | 0.00 | 73.22 | 73.22 | 27.11 | 46.11 |
| S4 | Area1 | -466.18 | 566.68 | 100.50 | / | 100.50 |
| | Area2 | 535.58 | -535.58 | 0.00 | / | 0.00 |
| | $s_{1\text{-}2}$ | 0.00 | -31.10 | -31.10 | 0.00 | -31.10 |
| S5 | Area1 | -463.58 | 566.60 | 103.01 | / | 103.01 |
| | Area2 | 556.88 | -556.88 | 0.00 | / | 0.00 |
| | $s_{1\text{-}2}$ | 0.00 | -9.72 | -9.72 | 22.16 | -31.88 |

### *C. Case 3: Three-Area (28 Nodes) Power System*

Next, we extend pricing and settlement approaches to a three-area power system with more interfaces. The system is composed of IEEE 14, 9, and 5-bus systems. We upload our MATLAB case file onto Github in [29]. The detailed network and the common equivalent boundary system are shown in Fig. 14. In this case, prices of interface bids are set as 0.1 $/MWh. Capacities of the congested internal line (1-2) and tie-line (10-6) are set as 5 MW.

Settlements of areas 1, 2, 3 and interface bids 1-2, 1-3, 2-3 are shown in Fig. 15. In this case, the congestion rent of tie-line (10-6) would be covered by interface bids 1-2, 1-3, 2-3. Congestion rent of internal line (1-2) in area 1 would be covered by area 1 and interface bids 1-2, 1-3, 2-3 together. Settlement results are consistent with the example in Fig. 8. Merchandise surplus of area 1 is equal to congestion rent afforded by area 1, and results of area 2 and area 3 are similar. The profit of interface bids 1-2 is equal to congestion rent afforded by them. And results of interface bids 1-3 and 2-3 are similar.

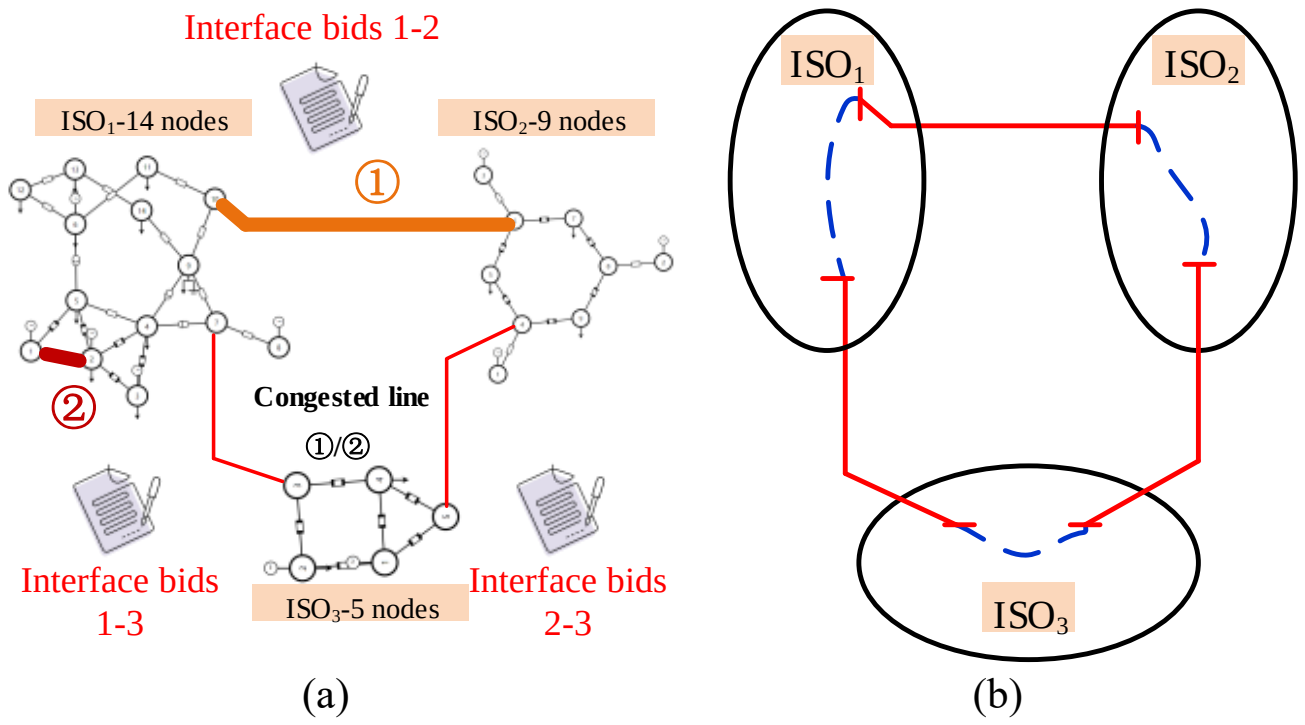

Fig. 14. Three-area power system in case 3 (a) original network (b) common boundary system after doing network equivalence

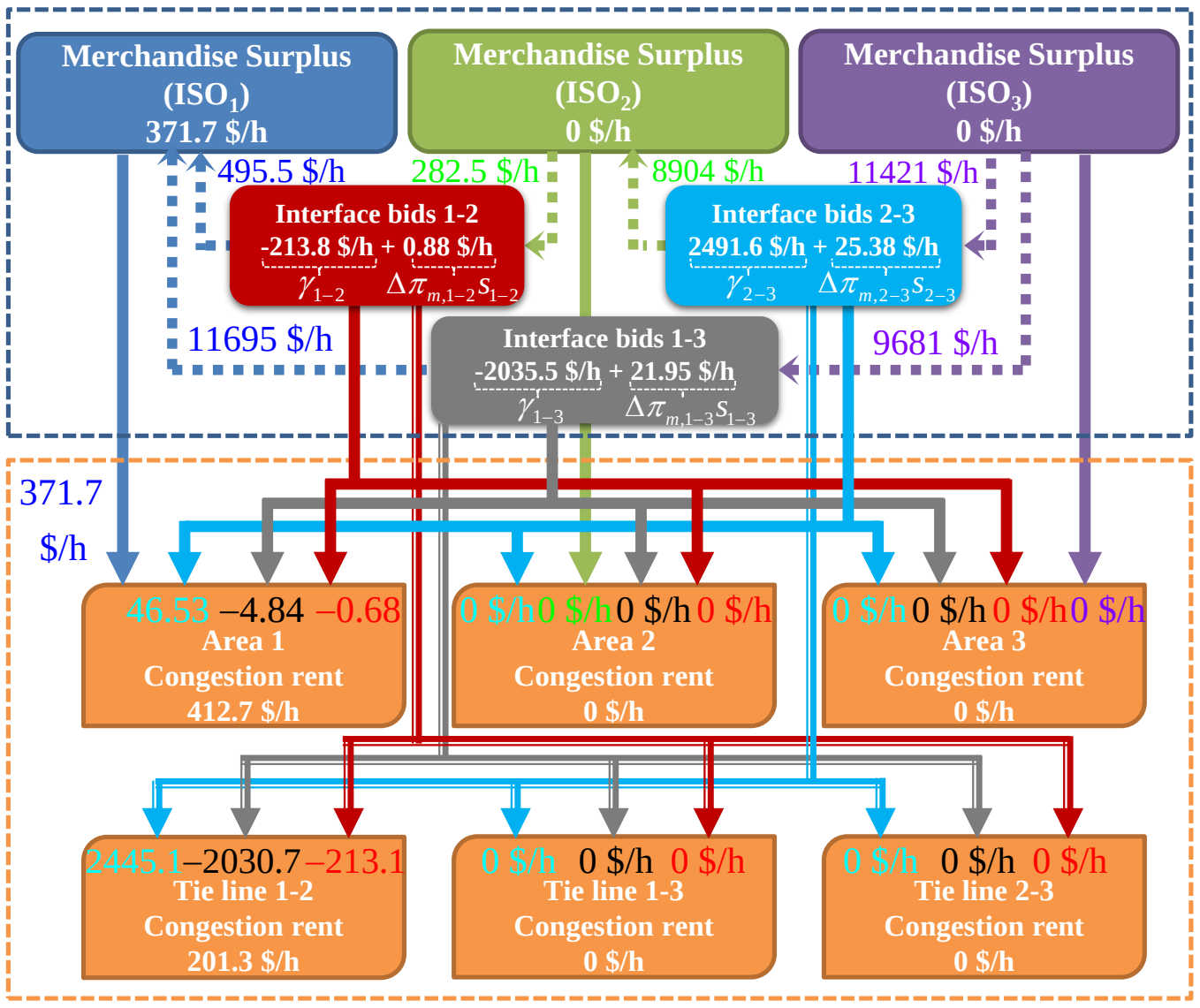

Fig. 15. Results of jointly covering congestion rents in case 3

## VII. Conclusion

In this paper, approaches of joint covering congestion rents in the multi-area power system considering loop flow effects are illustrated. We develop the framework of quantifying and pricing the contribution of each area and interface bids to line congestions. When settling a multi-area power system, we develop a distributed LMP recovery method to protect the privacy of each area. Under the assumption of the DCOPF model and no degeneracy, calculation results of distributed LMP recovery are equal to that of the centralized GCTS model. Finally, the contribution of each area and interface bids to line congestions is quantified and priced based on shift factors and shadow prices. And congestion rents are jointly covered among all areas and interface bids considering loop flow effects.

In the future, pricing and settlement approaches will be studied further in the line expansion to improve long-term efficiency.

## Appendix

### A. Proof of Theorem 1

The key issue of recovering global LMPs is determining the marginal units under the pre-determined scheduling of generators and interface bids. To simplify the explanation, the GCTS model of (2) ~ (7) is reformulated as follows:

$$\min_{\boldsymbol{g},\boldsymbol{s}} \; C = \boldsymbol{c}^T\boldsymbol{g} + \Delta\boldsymbol{\pi}^T\boldsymbol{s}\,, \tag{29}$$

$$h(\boldsymbol{g},\boldsymbol{s}) = 0 \qquad \boldsymbol{\lambda}, \tag{30}$$

$$f_l^L(\boldsymbol{g},\boldsymbol{s}) \le 0 \qquad \boldsymbol{\mu}, \tag{31}$$

$$g_g^G(\boldsymbol{g}) \le 0 \qquad \boldsymbol{\eta}_g, \tag{32}$$

$$g_s^S(\boldsymbol{s}) \le 0 \qquad \boldsymbol{\eta}_s, \tag{33}$$

where (30) represents power balance constraint, (31) represents line capacity limits, and (32), (33) represent the capacity limits of generators and interface bids. Multipliers of all constraints are listed on the right column.

After solving the GCTS model, binding constraints in (31), (32), (33) correspond to congested lines, non-marginal generators, and non-marginal interface bids, respectively. Those three sets are shown as:

$$A_L(\boldsymbol{g}^*,\boldsymbol{s}^*) = \left\{ l : 1 \le l \le N_L, g_l^L(\boldsymbol{g}^*,\boldsymbol{s}^*) = 0 \right\}, \tag{34}$$

$$A_G(\boldsymbol{g}^*,\boldsymbol{s}^*) = \left\{ g : 1 \le g \le N_G, g_g^G(\boldsymbol{g}^*,\boldsymbol{s}^*) = 0 \right\}, \tag{35}$$

$$A_S(\boldsymbol{g}^*,\boldsymbol{s}^*) = \left\{ s : 1 \le s \le N_S, g_s^D(\boldsymbol{g}^*,\boldsymbol{s}^*) = 0 \right\}, \tag{36}$$

where $N_L$, $N_G$, $N_S$ represent the number of congested lines, non-marginal generators, and non-marginal interface bids. Vector $\boldsymbol{g}^*$, $\boldsymbol{s}^*$ represent the optimal primal solution of the GCTS model.

With no degeneracy, market-clearing results satisfy the linear independence constraint qualification (LICQ) condition [30], which means the row vectors of gradient matrix $\boldsymbol{D}$ in (37) are linear independent.

$$\begin{aligned}\boldsymbol{D} = \nabla h(\boldsymbol{g}^*,\boldsymbol{s}^*) &\cup \left\{ \nabla g_l^L(\boldsymbol{g}^*,\boldsymbol{s}^*) \right\}_{l\in A_L(\boldsymbol{g}^*,\boldsymbol{s}^*)} \\ &\cup \left\{ \nabla g_g^G(\boldsymbol{g}^*,\boldsymbol{s}^*) \right\}_{l\in A_G(\boldsymbol{g}^*,\boldsymbol{s}^*)} \\ &\cup \left\{ \nabla g_s^S(\boldsymbol{g}^*,\boldsymbol{s}^*) \right\}_{l\in A_S(\boldsymbol{g}^*,\boldsymbol{s}^*)}\end{aligned} \tag{37}$$

After performing the elementary transformation, the gradient matrix will be obtained by

$$\boldsymbol{D}' = \begin{bmatrix} \boldsymbol{1}_{1\times N_m} & \boldsymbol{0} \\ \boldsymbol{S}_{N_L\times N_m} & \boldsymbol{0} \\ \boldsymbol{0} & \boldsymbol{I} \end{bmatrix} = \begin{bmatrix} \dot{\boldsymbol{D}}_{(1+N_L)\times N_m} & \boldsymbol{0} \\ \boldsymbol{0} & \boldsymbol{I} \end{bmatrix}, \tag{38}$$

where $N_m$ represents the number of marginal units, vector $\boldsymbol{S}_{N_L\times N_m}$ represents the shift factor between congested lines and marginal units. Matrix $\dot{\boldsymbol{D}}_{(1+N_L)\times N_m}$ is corresponding to all congested lines and marginal units, which is used to calculate non-zero Lagrange multipliers.

Hence, first-order optimality conditions associated with marginal units are sufficient to calculate LMP under the assumption of DCOPF and no degeneracy. Thus LMP recovery results are equal to LMPs obtained from the GCTS model. □

## B. Proof of Theorem 2

Due to network equivalence, we further derive the right side of (22) as

$$
\begin{aligned}
\boldsymbol{\mu}_{i-j}^T \boldsymbol{f}_{i-j} =& \boldsymbol{\mu}_{i-j}^T \big(-\boldsymbol{S}_{i-j,Bi}\boldsymbol{B}_{\bar{i}i}\boldsymbol{B}_{ii}^{-1}(\boldsymbol{g}_i-\boldsymbol{d}_i) \\
&-\boldsymbol{S}_{i-j,Bm}\boldsymbol{B}_{\bar{m}m}\boldsymbol{B}_{mm}^{-1}(\boldsymbol{g}_m-\boldsymbol{d}_m) \\
&-\boldsymbol{S}_{i-j,Bj}\boldsymbol{B}_{\bar{j}j}\boldsymbol{B}_{jj}^{-1}(\boldsymbol{g}_j-\boldsymbol{d}_j)\big) \\
=& \boldsymbol{\mu}_{i-j}^T\left(\begin{bmatrix}\boldsymbol{S}_{i-j,Bi} & \boldsymbol{S}_{i-j,Bm} & \boldsymbol{S}_{i-j,Bj}\end{bmatrix}\boldsymbol{Ms}\right) \\
=& \boldsymbol{1}^T\boldsymbol{\psi}_{i-j,i-j}+\boldsymbol{1}^T\boldsymbol{\psi}_{i-j,i-m}+\boldsymbol{1}^T\boldsymbol{\psi}_{i-j,j-m}=\beta_{i-j},
\end{aligned} \tag{39}
$$

$$
\begin{aligned}
\boldsymbol{\mu}_i^T \boldsymbol{f}_i =& \boldsymbol{\mu}_i^T\big(\boldsymbol{S}_{i,i}(\boldsymbol{g}_i-\boldsymbol{d}_i)-\boldsymbol{S}_{i,Bm}\boldsymbol{B}_{\bar{m}m}\boldsymbol{B}_{mm}^{-1}(\boldsymbol{g}_m-\boldsymbol{d}_m) \\
&-\boldsymbol{S}_{i,Bj}\boldsymbol{B}_{\bar{j}j}\boldsymbol{B}_{jj}^{-1}(\boldsymbol{g}_j-\boldsymbol{d}_j)\big) \\
=& \boldsymbol{\mu}_i^T\big(\boldsymbol{S}_{i,i}(\boldsymbol{g}_i-\boldsymbol{d}_i)-\boldsymbol{S}_{i,Bi}\boldsymbol{B}_{\bar{i}i}\boldsymbol{B}_{ii}^{-1}(\boldsymbol{g}_i-\boldsymbol{d}_i) \\
&-\boldsymbol{S}_{i,Bm}\boldsymbol{B}_{\bar{m}m}\boldsymbol{B}_{mm}^{-1}(\boldsymbol{g}_m-\boldsymbol{d}_m)-\boldsymbol{S}_{i,Bj}\boldsymbol{B}_{\bar{j}j}\boldsymbol{B}_{jj}^{-1}(\boldsymbol{g}_j-\boldsymbol{d}_j)\big) \\
=& \boldsymbol{\mu}_i^T\left(\boldsymbol{S}_{i,i}(\boldsymbol{g}_i-\boldsymbol{d}_i)+\begin{bmatrix}\boldsymbol{S}_{i,Bi} & \boldsymbol{S}_{i,Bm} & \boldsymbol{S}_{i,Bj}\end{bmatrix}\boldsymbol{Ms}\right) \\
=& \psi_{i,i}+\boldsymbol{1}^T\boldsymbol{\psi}_{i,i-j}+\boldsymbol{1}^T\boldsymbol{\psi}_{i,i-m}+\boldsymbol{1}^T\boldsymbol{\psi}_{i,j-m}=\beta_i.
\end{aligned} \tag{40}
$$

Thus, theorem 2 is verified. □

## C. Proof of Theorem 3

The merchandise surplus of the area $i$ $MS_i$ is derived as

$$
\begin{aligned}
MS_i =& \nabla_{\boldsymbol{d}_i}^T C^*(\boldsymbol{d}_i-\boldsymbol{g}_i)+\nabla_{\boldsymbol{M}_i\boldsymbol{s}}^T C^*\times\boldsymbol{M}_i\boldsymbol{s} \\
=& \left(\lambda\times\boldsymbol{1}-\boldsymbol{S}_{i,i}^T\boldsymbol{\mu}_i-\boldsymbol{S}_{t,i}^T\boldsymbol{\mu}_t+(\boldsymbol{B}_{\bar{i}i}\boldsymbol{B}_{ii}^{-1})^T\boldsymbol{\rho}_i\right)^T(\boldsymbol{d}_i-\boldsymbol{g}_i) \\
&+\left(\lambda\times\boldsymbol{1}-\boldsymbol{S}_{i,Bi}^T\boldsymbol{\mu}_i-\boldsymbol{S}_{t,Bi}^T\boldsymbol{\mu}_t+\boldsymbol{\rho}_i\right)^T\boldsymbol{M}_i\boldsymbol{s} \\
=& -\boldsymbol{\mu}_i^T\boldsymbol{S}_{i,i}(\boldsymbol{d}_i-\boldsymbol{g}_i)-\boldsymbol{\mu}_t^T\boldsymbol{S}_{t,i}(\boldsymbol{d}_i-\boldsymbol{g}_i) \\
&+\boldsymbol{\mu}_i^T\boldsymbol{S}_{i,Bi}\boldsymbol{B}_{\bar{i}i}\boldsymbol{B}_{ii}^{-1}(\boldsymbol{d}_i-\boldsymbol{g}_i)+\boldsymbol{\mu}_t^T\boldsymbol{S}_{t,Bi}\boldsymbol{B}_{\bar{i}i}\boldsymbol{B}_{ii}^{-1}(\boldsymbol{d}_i-\boldsymbol{g}_i) \\
=& \boldsymbol{\mu}_i^T\boldsymbol{S}_{i,i}(\boldsymbol{g}_i-\boldsymbol{d}_i)=\gamma_i.
\end{aligned} \tag{41}
$$

The arbitrage revenue of a given interface bid buying from area $j$ and selling to the area $i$ $R_{i\text{-}j}$ is derived as

$$
\begin{aligned}
R_{i-j} =& \nabla_{\boldsymbol{M}_i s_{i-j}}^T C^*\times\boldsymbol{M}_i s_{i-j}-\nabla_{\boldsymbol{M}_j s_{i-j}}^T C^*\times\boldsymbol{M}_j s_{i-j} \\
=& (\lambda\times\boldsymbol{1}-\boldsymbol{S}_{i,Bi}^T\boldsymbol{\mu}_i-\boldsymbol{S}_{t,Bi}^T\boldsymbol{\mu}_t-\boldsymbol{S}_{j,Bi}^T\boldsymbol{\mu}_j \\
&-\boldsymbol{S}_{m,Bi}^T\boldsymbol{\mu}_m+\boldsymbol{\rho}_i)^T\boldsymbol{M}_i s_{i-j}-(\lambda\times\boldsymbol{1}-\boldsymbol{S}_{j,Bj}^T\boldsymbol{\mu}_j \\
&-\boldsymbol{S}_{t,Bj}^T\boldsymbol{\mu}_t-\boldsymbol{S}_{i,Bj}^T\boldsymbol{\mu}_i-\boldsymbol{S}_{m,Bj}^T\boldsymbol{\mu}_m+\boldsymbol{\rho}_j)^T\boldsymbol{M}_j s_{i-j} \\
=& \Delta\pi_{m,i-j}s_{i-j}+\boldsymbol{\mu}_i^T(\boldsymbol{S}_{i,Bj}-\boldsymbol{S}_{i,Bi})s_{i-j} \\
&+\boldsymbol{\mu}_j^T(\boldsymbol{S}_{j,Bj}-\boldsymbol{S}_{j,Bi})s_{i-j}+\boldsymbol{\mu}_m^T(\boldsymbol{S}_{m,Bj}-\boldsymbol{S}_{m,Bi})s_{i-j} \\
&+\boldsymbol{\mu}_t^T(\boldsymbol{S}_{t,Bj}-\boldsymbol{S}_{t,Bi})s_{i-j} \\
=& \Delta\pi_{m,i-j}s_{i-j}+\psi_{i,i-j}+\psi_{j,i-j}+\psi_{m,i-j} \\
&+\psi_{i-j,i-j}+\psi_{i-m,i-j}+\psi_{j-m,i-j}=\Delta\pi_{m,i-j}s_{i-j}+\gamma_{i-j}.
\end{aligned} \tag{42}
$$

Thus, theorem 3 is verified. □

## D. Proof of Theorem 4

According to [22], the sensitivity of the local optimal cost of area $i$ with respect to cleared quantities of interface bids $\boldsymbol{s}$ is calculated as

$$
\nabla_{\boldsymbol{s}} C_i^* = \boldsymbol{M}^T\begin{bmatrix}\tilde{\boldsymbol{B}}_{\bar{i}\bar{i}} & \boldsymbol{B}_{\bar{i}\bar{j}} \\ \boldsymbol{B}_{\bar{j}\bar{i}} & \tilde{\boldsymbol{B}}_{\bar{j}\bar{j}}\end{bmatrix}^{-1}\left(\begin{bmatrix}\boldsymbol{B}_{\bar{i}i} & \boldsymbol{B}_{\bar{i}\bar{i}} \\ & \boldsymbol{B}_{\bar{j}\bar{i}}\end{bmatrix}\begin{bmatrix}\boldsymbol{\xi}_i \\ \bar{\boldsymbol{\xi}}_i\end{bmatrix}+\begin{bmatrix}\boldsymbol{H}_{\bar{i}}^T\boldsymbol{\mu}_i \\ \boldsymbol{0}\end{bmatrix}\right), \tag{43}
$$

where $\boldsymbol{\xi}_i$, $\bar{\boldsymbol{\xi}}_i$ represent the LMPs of internal and boundary nodes of area $i$, vector $\boldsymbol{B}_{\bar{i}\bar{j}}$ represents the sub-matrix between boundary buses in areas $i$ and $j$, vector $\tilde{\boldsymbol{B}}_{\bar{i}\bar{i}}$ represents the equivalent self-susceptance of boundary buses of area $i$. Other terms in (43) are similarly defined.

For area $i$, from the optimality condition of the local economic dispatch model in [22], we have

$$
\boldsymbol{B}_{\bar{i}i}\boldsymbol{\xi}_i+\boldsymbol{B}_{\bar{i}\bar{i}}\bar{\boldsymbol{\xi}}_i+\boldsymbol{H}_{\bar{i}}^T\boldsymbol{\mu}_i=0. \tag{44}
$$

By substituting (43) and (44) into the left side of (28), we have

$$
\begin{aligned}
&\nabla_{\boldsymbol{s}}{}^T C_i^*-\boldsymbol{\mu}_i^T\boldsymbol{S}_{i,B}\boldsymbol{M} \\
&=\left(\begin{bmatrix}\boldsymbol{B}_{\bar{i}i} & \boldsymbol{B}_{\bar{i}\bar{i}} \\ & \boldsymbol{B}_{\bar{j}\bar{i}}\end{bmatrix}\begin{bmatrix}\boldsymbol{\xi}_i \\ \bar{\boldsymbol{\xi}}_i\end{bmatrix}+\begin{bmatrix}\boldsymbol{H}_{\bar{i}}^T\boldsymbol{\mu}_i \\ \boldsymbol{0}\end{bmatrix}\right)^T\begin{bmatrix}\tilde{\boldsymbol{B}}_{\bar{i}\bar{i}} & \boldsymbol{B}_{\bar{i}\bar{j}} \\ \boldsymbol{B}_{\bar{j}\bar{i}} & \tilde{\boldsymbol{B}}_{\bar{j}\bar{j}}\end{bmatrix}^{-1}\boldsymbol{M} \\
&\quad-\boldsymbol{\mu}_i^T\boldsymbol{S}_{i,B}\boldsymbol{M} \\
&=\left(\begin{bmatrix}\tilde{\boldsymbol{B}}_{\bar{i}\bar{i}}\bar{\boldsymbol{\xi}}_i-\tilde{\boldsymbol{H}}_{\bar{i}}^T\boldsymbol{\mu}_i \\ \boldsymbol{B}_{\bar{j}\bar{i}}\bar{\boldsymbol{\xi}}_i\end{bmatrix}^T-\begin{bmatrix}\boldsymbol{\mu}_i^T\tilde{\boldsymbol{H}} \\ \boldsymbol{0}\end{bmatrix}\right)\begin{bmatrix}\tilde{\boldsymbol{B}}_{\bar{i}\bar{i}} & \boldsymbol{B}_{\bar{i}\bar{j}} \\ \boldsymbol{B}_{\bar{j}\bar{i}} & \tilde{\boldsymbol{B}}_{\bar{j}\bar{j}}\end{bmatrix}^{-1}\boldsymbol{M} \\
&=\begin{bmatrix}\bar{\boldsymbol{\xi}}_i^T & \boldsymbol{0}\end{bmatrix}\begin{bmatrix}\tilde{\boldsymbol{B}}_{\bar{i}\bar{i}} & \boldsymbol{B}_{\bar{i}\bar{j}} \\ \boldsymbol{B}_{\bar{j}\bar{i}} & \tilde{\boldsymbol{B}}_{\bar{j}\bar{j}}\end{bmatrix}\begin{bmatrix}\tilde{\boldsymbol{B}}_{\bar{i}\bar{i}} & \boldsymbol{B}_{\bar{i}\bar{j}} \\ \boldsymbol{B}_{\bar{j}\bar{i}} & \tilde{\boldsymbol{B}}_{\bar{j}\bar{j}}\end{bmatrix}^{-1}\begin{bmatrix}\boldsymbol{M}_i \\ \boldsymbol{M}_j\end{bmatrix} \\
&=\bar{\boldsymbol{\xi}}_i^T\boldsymbol{M}_i=\nabla_{\boldsymbol{M}_i\boldsymbol{s}}^T C^*\times\boldsymbol{M}_i.
\end{aligned} \tag{45}
$$

□

# References

[1] Williams, James H., et al. "Carbon‐neutral pathways for the United States." *AGU Advances* 2.1 (2021): e2020AV000284.

[2] Gössling, Stefan. "Carbon neutral destinations: A conceptual analysis." *Journal of Sustainable Tourism* 17.1 (2009): 17-37.

[3] Olhoff, Anne, and John M. Christensen. "Emissions Gap Report 2020." (2020). https://reliefweb.int/sites/reliefweb.int/files/resources/EGR20.pdf

[4] Ali, S. M., et al. "Wide area smart grid architectural model and control: A survey." *Renewable and Sustainable Energy Reviews* 64 (2016): 311-328.

[5] M. White and R. Pike, "ISO New England and New York ISO interregional interchange scheduling: Analysis and options," Jan. 2011, [Online]. Available: https://www.isone.com/staticassets/documents/pubs/whtpprs/iris_white_paper.pdf

[6] "FERC approves coordinated transaction scheduling for PJM and NYISO," March 2014, [Online]. Available: http://www.pjm.com/~/media/about-pjm/newsroom/2014-releases/20140313-coordinated-transaction-scheduling-PJM-NYISO.ashx.

[7] M. Liu and G. Gross, "Role of distribution factors in congestion revenue rights applications," in *IEEE Transactions on Power Systems*, vol. 19, no. 2, pp. 802-810, May 2004.

[8] H. Ming and L. Xie, "Revenue adequacy of wholesale electricity markets with demand response providers," *2016 IEEE Power and Energy Society General Meeting (PESGM)*, Boston, MA, 2016, pp. 1-5.

[9] U.S. Energy Information Administration, "Loop flow," [Online]. Available: https://www.eia.gov/tools/glossary/?id=l

[10] "New York Independent System Operator, Lake Erie Loop Flow Mitigation," November 2008, [Online]. Available: https://studylib.net/doc/18746699/lake-erie-loop-flow-mitigation

[11] "Clean Energy Wire, Loop flows: Why is wind power from northern Germany putting east European grids under pressure?" December 2015, [Online]. Available: https://www.cleanenergywire.org/factsheets/loop-

flows-why-wind-power-northern-germany-putting-east-european-grids-under-pressure
[12] LI Xiang, et al. "Analysis of AC Section Power Fluctuation in West-to-East Electricity Transmission Project and Related Suggestion," in *Southern Power System Technology*, vol. 5, pp. 18-22, Dec. 2011.
[13] RTO Insider, SPP to Run Congestion Plan for CAISO, Others, August 2018, http://www.rtoinsider.com/spp-wiufmp-congestion-controllable-devices-98587/
[14] THEMA Consulting Group, Loop flows--final advice, Prepared for The European Commissions, October 2013
[15] A. J. Conejo and J. A. Aguado, "Multi-area coordinated decentralized DC optimal power flow," in *IEEE Transactions on Power Systems*, vol. 13, no. 4, pp. 1272-1278, Nov. 1998.
[16] Zhao W, Liu M, Zhu J, et al. Fully decentralised multi-area dynamic economic dispatch for large-scale power systems via cutting plane consensus[J]. *IET Generation, Transmission & Distribution*, 2016, 10(10): 2486-2495.
[17] F. Zhao, E. Litvinov and T. Zheng, "A Marginal Equivalent Decomposition Method and Its Application to Multi-Area Optimal Power Flow Problems," in *IEEE Transactions on Power Systems*, vol. 29, no. 1, pp. 53-61, Jan. 2014.
[18] Y. Guo, L. Tong, W. Wu, B. Zhang and H. Sun, "Coordinated Multi-Area Economic Dispatch via Critical Region Projection," in *IEEE Transactions on Power Systems*, vol. 32, no. 5, pp. 3736-3746, Sept. 2017.
[19] Y. Guo, S. Bose and L. Tong, "On Robust Tie-Line Scheduling in Multi-Area Power Systems," in *IEEE Transactions on Power Systems*, vol. 33, no. 4, pp. 4144-4154, July 2018.
[20] Y. Ji and L. Tong, "Multi-Area Interchange Scheduling Under Uncertainty," in *IEEE Transactions on Power Systems*, vol. 33, no. 2, pp. 1659-1669, March 2018.
[21] C. K. Hogan, B. Kranz, D. Laplante, et al. "I. Proxy Buses and Congestion Pricing of Inter-Balancing Authority Area Transactions," 2008, [Online]. Available: http://lmpmarketdesign.com/papers/Proxy_Buses_and_Congestion_Pricing_06_09_08.pdf
[22] Y. Guo, Y. Ji and L. Tong, "Generalized Coordinated Transaction Scheduling: A Market Approach to Seamless Interfaces," in *IEEE Transactions on Power Systems*, vol. 33, no. 5, pp. 4683-4693, Sept. 2018.
[23] Y. Ji, T. Zheng and L. Tong, "Stochastic Interchange Scheduling in the Real-Time Electricity Market," in *IEEE Transactions on Power Systems*, vol. 32, no. 3, pp. 2017-2027, May 2017
[24] L. Zhang et al., "Congestion Surplus Minimization Pricing Solutions When Lagrange Multipliers are not Unique," in *IEEE Transactions on Power Systems*, vol. 29, no. 5, pp. 2023-2032, Sept. 2014.
[25] D. Feng, Z. Xu, J. Zhong and J. Ostergaard, "Spot Pricing When Lagrange Multipliers Are Not Unique," in *IEEE Transactions on Power Systems*, vol. 27, no. 1, pp. 314-322, Feb. 2012.
[26] Wang P, Mou S, Lian J, et al. "Solving a system of linear equations: From centralized to distributed algorithms," in *Annual Reviews in Control*, vol. 47, pp. 306-322, 2019.
[27] S. Mou, J. Liu and A. S. Morse, "A Distributed Algorithm for Solving a Linear Algebraic Equation," in *IEEE Transactions on Automatic Control*, vol. 60, no. 11, pp. 2863-2878, Nov. 2015.
[28] Oren, Shmuel S., and Kory W. Hedman. "Revenue adequacy, shortfall allocation and transmission performance incentives in FTR/FGR markets." *2010 IREP Symposium Bulk Power System Dynamics and Control-VIII (IREP)*. IEEE, 2010.
[29] H. Liu, “Modified three area case.”, https://github.com/thuliuhao/Modified-three-area-case
[30] D. Feng, C. Liu, Z. Liu, L. Zhang, Y. Ding and C. Campaigne, "Constraint Qualification Based Detection Method for Nodal Price Multiplicity," in *IEEE Transactions on Power Systems*, vol. 32, no. 6, pp. 4968-4969, Nov. 2017.